\documentclass{article}

\usepackage[colorlinks = true,
            linkcolor = blue,
            urlcolor  = blue,
            citecolor = blue,
            anchorcolor = blue]{hyperref}
\usepackage{microtype}
\usepackage{graphicx}
\usepackage[font=small,labelfont=bf]{caption}
\usepackage{subfigure}
\usepackage{booktabs} 
\usepackage{enumitem}
\usepackage{bbm}
\usepackage{subfigure}
\usepackage{float}
\usepackage{amsmath, amssymb, mathtools,amsthm}
\usepackage{enumitem}
\usepackage{url}
\usepackage[capitalize,noabbrev]{cleveref}
\usepackage{pifont}
\newcommand{\cmark}{\ding{51}}%
\newcommand{\xmark}{\ding{55}}%
\newcommand{\vequalsignnospace}{\texttt{=}}

\usepackage[accepted]{icml2021}

\begin{document}

\twocolumn[

\icmltitle{AudioLDM: Text-to-Audio Generation with Latent Diffusion Models}




\icmlsetsymbol{equal}{*}

\begin{icmlauthorlist}
\icmlauthor{Haohe Liu}{equal,su}
\icmlauthor{Zehua Chen}{equal,im}
\icmlauthor{Yi Yuan}{su}
\icmlauthor{Xinhao Mei}{su} 
\icmlauthor{Xubo Liu}{su}
\icmlauthor{Danilo Mandic}{im} \\
\icmlauthor{Wenwu Wang}{su} 
\icmlauthor{Mark D. Plumbley}{su}

\end{icmlauthorlist}

\icmlaffiliation{su}{Centre for Vision, Speech and Signal Processing (CVSSP), University of Surrey, Guildford, UK}
\icmlaffiliation{im}{Department of Electrical and Electronic Engineering, Imperial College London, London, UK}
\icmlcorrespondingauthor{Haohe Liu}{haohe.liu@surrey.ac.uk}

\icmlkeywords{Machine Learning, ICML}

\vskip 0.3in
]



\printAffiliationsAndNotice{\icmlEqualContribution} 


\begin{abstract}

%

Text-to-audio (TTA) systems have recently gained attention for their ability to synthesize general audio based on text descriptions.
However, previous studies in TTA have limited generation quality with high computational costs.
In this study, we propose AudioLDM, a TTA system that is built on a latent space to learn continuous audio representations from contrastive language-audio pretraining (CLAP) embeddings.
The pretrained CLAP models enable us to train LDMs with audio embeddings while providing text embeddings as the condition during sampling. 
By learning the latent representations of audio signals without modelling the cross-modal relationship, AudioLDM improves both generation quality and computational efficiency. 
Trained on AudioCaps with a single GPU, AudioLDM achieves state-of-the-art TTA performance compared to other open-sourced systems, measured by both objective and subjective metrics. AudioLDM is also the first TTA system that enables various text-guided audio manipulations~(e.g., style transfer) in a zero-shot fashion. Our implementation and demos are available at \url{https://audioldm.github.io}.
\end{abstract}


\section{Introduction}
\label{Introduction}


Generating sound effects, music, or speech according to personalized requirements is important for applications such as augmented and virtual reality, game development, and video editing. Traditionally, audio generation has been achieved through signal processing techniques~\cite{andresen1979new, karplus1983digital}. In recent years, generative models~\cite{oord2016wavenet, DDPM, SGM, tan2022naturalspeech}, either unconditional or conditioned on other modalities~\cite{kreuk2022audiogen, zelaszczyk2022audio}, have revolutionized this task. Previous studies primarily worked on the label-to-sound setting with a small set of labels~\cite{liu2021conditional,pascual2022full} such as the ten sound classes in the UrbanSound8K dataset~\cite{salamon2014dataset}.
In comparison, natural language is considerably more flexible than labels as they can include fine-grained descriptions of audio signals, such as pitch, acoustic environment, and temporal order. 
The task of generating audio prompted with natural language descriptions is known as \textit{text-to-audio}~(TTA) generation.

TTA systems are designed to generate a wide range of high-dimensional audio signals. To efficiently model the data, we adopt a similar approach as DiffSound~\cite{yang2022diffsound} by employing a learned discrete representation to efficiently model high-dimensional audio signals. We also draw inspiration from the recent advancements in autoregressive modelling of discrete representation learnt on the waveform, such as AudioGen~\cite{kreuk2022audiogen}, which has surpassed the capabilities of DiffSound. Building on the success of StableDiffusion~\cite{rombach2022high}, which uses latent diffusion models (LDMs) for high-quality image generation, we extend previous TTA approaches to continuous latent representations, instead of learning discrete representations. Additionally, as audio manipulations, such as style transfer~\cite{engel2020ddsp, pascual2022full}, are desired for some applications such as games, 
we explore and achieve various zero-shot text-guided audio manipulations with LDMs, which have not been demonstrated before.

\begin{figure*}
    \centerline{
    \includegraphics[width=\linewidth]{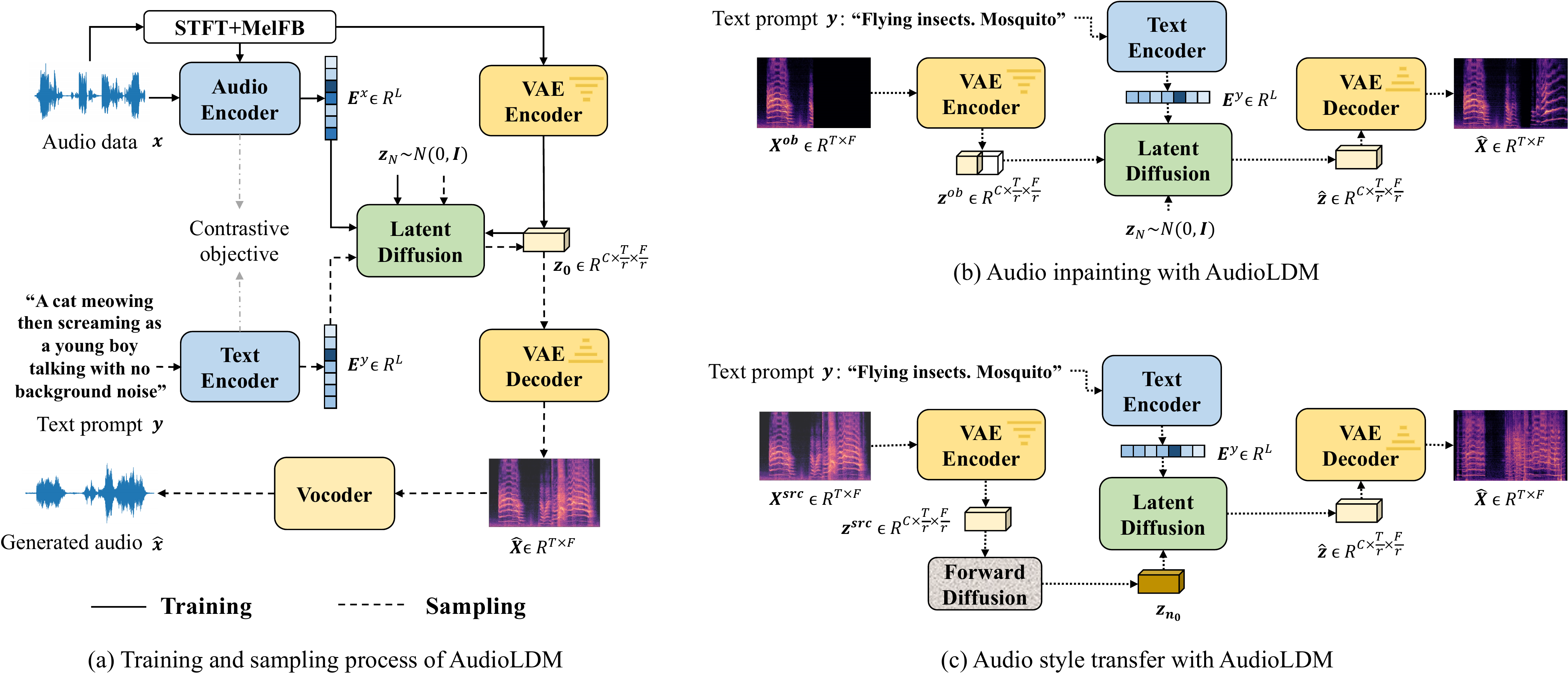}}
    \caption{Overview of the AudioLDM system for text-to-audio generation~(a). During training, latent diffusion models~(LDMs) are conditioned on an audio embedding $\boldsymbol{E}^{x}$ and trained in a continuous space $\boldsymbol{z}_{0}$ learned by VAE. The sampling process uses text embedding $\boldsymbol{E}^{y}$ as the condition. Given a pretrained LDM, zero-shot audio inpainting~(b) and style transfer~(c) are realized in the reverse diffusion process of LDM. The block \textit{Forward Diffusion} denotes the process that corrupt data with gaussian noise~(see Equation~\ref{forwardprocess}).}
    \label{fig:overalldesign}
\end{figure*}

For previous TTA works, a potential limitation for generation quality is the requirement of large-scale high-quality audio-text data pairs, which are usually not readily available, and where they are available, are  
of limited quality and quantity~\cite{liu2022separate}. To better utilize the low-quality data, several methods for text preprocessing have been proposed~\cite{kreuk2022audiogen, yang2022diffsound}. However, these preprocessing steps limit generation performances by overlooking the relations of sound events~
(e.g., \textit{a dog is barking at the bark} is transformed into \textit{dog bark park}).
By comparison, our proposed method only requires audio data for generative model training, circumvents the challenge of text preprocessing, and performs better than using audio-text paired data, as we will discuss later.

In this work, we present a TTA system, AudioLDM, which achieves high generation quality with continuous LDMs, with good computational efficiency and enables text-conditional audio manipulations.
The overview of AudioLDM design for TTA generation and text-guided audio manipulation is shown in Figure~\ref{fig:overalldesign}. 
Specifically, AudioLDM learns to generate the representation in a latent space encoded by a mel-spectrogram-based variational auto-encoder~(VAE). An LDM conditioned on a contrastive language-audio pretraining~(CLAP) embedding is developed for VAE latent generation. By leveraging the audio-text-aligned embedding space in CLAP, we remove the requirement for paired audio-text data during training LDM, as the condition for VAE latent generation can directly come from the audio itself. We demonstrate that training an LDM with audio only can be even better than training with audio-text data pairs. The proposed AudioLDM achieves leading TTA performance on the AudioCaps dataset with a Freshet distance~(FD) of $23.31$, outperforming the DiffSound baseline~(FD of $47.68$) by a large margin. Our system also enables zero-shot audio manipulations in the sampling process. In summary, our contributions are as follows:
\begin{list}{\labelitemi}{\leftmargin=1em}
    \setlength{\itemsep}{1pt}
    \setlength{\parskip}{0pt}
    \item We demonstrate the first attempt to develop a continuous LDM for TTA generation. Our AudioLDM method outperforms existing methods in both subjective evaluation and objective metrics.
    \item We utilize CLAP embeddings to enable TTA generation without using language-audio pairs to train LDMs. 
    \item We experimentally show that using audio only data in LDM training can obtain a high-quality and computationally efficient TTA system. 
    \item We show that our proposed TTA system can perform text-guided audio styles manipulation, such as audio style transfer, super-resolution, and inpainting, without fine-tuning the model on a specific task.
\end{list}

\section{Related Work}
\label{Background}

\label{TTA}

\textbf{Text-to-Audio Generation}~has gained a lot of attention recently. Two works~\cite{yang2022diffsound,kreuk2022audiogen} explore how to learn audio representations in a discrete space given a natural language description, and then decode the representations to the audio waveform. Since both works require audio-text paired data for training the latent generation model, they have both proposed methods to address the issues of low quality and scarcity of paired data.

DiffSound~\cite{yang2022diffsound} consists of a text encoder, a decoder, a vector-quantized variational autoencoder~(VQ-VAE), and a vocoder. To alleviate the scarcity of audio-text paired data, they propose a mask-based text generation strategy~(MBTG) for generating text descriptions from audio labels. For example, the label \textit{dog bark, a man speaking} will be represented as \textit{[M] [M] dog bark [M] man speaking [M]}, where \textit{[M]} represent the mask token. However, the text generated by MBTG still only includes the label information, which might potentially limit model performance. 

AudioGen~\cite{kreuk2022audiogen} uses a Transformer-based decoder to learn to generate the target discrete tokens that are directly compressed from the waveform. AudioGen is trained on $10$ datasets and proposes data augmentation methods to enhance the diversity of training data. When creating the language-audio pairs, they pre-process the language descriptions to labels to better match the class-label annotation distribution and simplify the task. For example, 
the text description \textit{a dog is barking at the park} is transformed to \textit{dog bark park}. For data augmentation, they mix audio samples according to various signal-to-noise ratios and concatenate the transformed language descriptions. This means that the detailed text descriptions showing the spatial and temporal relationships are discarded.


\textbf{Diffusion Models}~\cite{DDPM, SGM} have achieved state-of-the-art sample quality in tasks such as image generation~\cite{DiffusionBeatsGANs, DALLE2, Imagen}, image restoration~\cite{ISRIR}, speech generation~\cite{WaveGrad, DiffWave, leng2022binauralgrad}, and video generation~\cite{MakeAVideo, ImagenVideo}. For speech or audio synthesis, diffusion models have been studied for both mel-spectrogram generation~\cite{Grad-TTS, ResGrad} and waveform generation~\cite{BDDM, PriorGrad, InferGrad}. 

A major concern with diffusion models is that the iterative generation process in a high-dimensional data space will result in a low inference speed. One of the solutions is to employ diffusion models in a small latent space, an approach used, for example, in image generation~\cite{LSGM, D2C, rombach2022high}. For TTA generation, the audio waveform has redundant information~\cite{liu2022simple, liu2022learning} that increases modeling complexity and decreases inference speed. To overcome this, DiffSound~\cite{yang2022diffsound} uses text-conditional discrete diffusion models to generate discrete tokens as a compressed representation of mel-spectrograms. However, the quality of the sound generated by their method is limited. In addition, audio manipulation methods are not explored. 


\section{Text-Conditional Audio Generation}
\label{AudioLDM}



\subsection{Contrastive Language-Audio Pretraining}
\label{CLAP}
Text-to-image generation models have shown stunning sample quality by utilizing Contrastive Language-Image Pretraining (CLIP)~\cite{CLIP} for generating the image prior. Inspired by this, we leverage Contrastive Language-Audio Pretraining~(CLAP)~\cite{wu2022large} to facilitate TTA generation.   

We denote audio samples as $x$ and the text description as $y$. A text encoder $f_{\text{text}}(\cdot)$ and an audio encoder $f_{\text{audio}}(\cdot)$ are used to extract a text embedding $\boldsymbol{E}^{y}\in \mathbb{R}^{L}$ and an audio embedding $\boldsymbol{E}^{x}\in \mathbb{R}^{L}$ respectively, where $L$ is the dimension of CLAP embedding. A recent study \cite{wu2022large} has explored different architectures for both the text encoder and the audio encoder when training the CLAP model. We follow their result to build an audio encoder based on HTSAT~\cite{HTSAT}, and built a text encoder based on RoBERTa~\cite{RoBERTa}. We use a symmetric cross-entropy loss as the training objective~\cite{CLIP,wu2022large}. For details of the training process and the language-audio datasets see Appendix~\ref{app:CLAP}. 

After training the CLAP model, an audio sample $x$ can be transformed into an embedding $\boldsymbol{E}^{x}$ within an aligned audio and text embedding space. The generalization ability of CLAP model has been demonstrated by various downstream tasks such as the zero-shot audio classification~\cite{wu2022large}. Then, for unseen language or audio samples, CLAP embeddings also provide cross-modal information.

\subsection{Conditional Latent Diffusion Models}
\label{CLDMs}

The TTA system can generate an audio sample $\hat{x}$ given text description $y$. With probabilistic generative model LDMs, we estimate the true conditional data distribution $q(\boldsymbol{z}_{0}|\boldsymbol{E}^{y})$ with a model distribution $p_{\theta}(\boldsymbol{z}_{0}|\boldsymbol{E}^{y})$, where $\boldsymbol{z}_{0}\in \mathbb{R}^{C\times \frac{T}{r}\times \frac{F}{r}}$ is the prior of an audio sample $x$ in the space formed from the compressed representation of the mel-spectrogram $\boldsymbol{X}\in \mathbb{R}^{T\times F}$, and $\boldsymbol{E}^{y}$ is the text embedding obtained by pretrained text encoder $f_{\text{text}}(\cdot)$ in CLAP. Here, $r$ denotes the compression level, $C$ denotes the channel of the compressed representation, $T$ and $F$ denote the time-frequency dimensions in the mel-spectrogram $\boldsymbol{X}$. With pretrained CLAP to jointly embed the audio and text information, the audio embedding $\boldsymbol{E}^{x}$ and the text embedding $\boldsymbol{E}^{y}$ share a joint cross-modal space. This allows us to provide $\boldsymbol{E}^{x}$ for training the LDMs, while providing $\boldsymbol{E}^{y}$ for TTA generation.

Diffusion models~\cite{DDPM, SGM} consist of two processes: i) a forward process to transform the data distribution into a standard Gaussian distribution with a predefined noise schedule $ 0 < \beta_{1} < \dots < \beta_{n} < \dots \beta_{N} < 1$, and ii) a reverse process to gradually generate data samples from the noise according to an inference noise schedule. 

In the forward process, at each time step $n\in [1,\dots,N]$, the transition probability is given by:
\begin{align}
q(\boldsymbol{z}_{n}|\boldsymbol{z}_{n-1})&=\mathcal{N}(\boldsymbol{z}_{n};\sqrt{1-\beta_{n}}\boldsymbol{z}_{n-1},\beta_{n}\boldsymbol{I}), \\
\label{forwardprocess}
q(\boldsymbol{z}_{n}|\boldsymbol{z}_{0})&=\mathcal  N(\boldsymbol{z}_{n};\sqrt{\bar{\alpha}_{n}}\boldsymbol{z}_{0},(1-\bar{\alpha}_{n})\boldsymbol{\epsilon}),
\end{align} 
where $\boldsymbol{\epsilon}\sim\mathcal N(\boldsymbol{0},\boldsymbol{I})$ denotes injected noise, $\alpha_{n}$ is a reparameterization of $1-\beta_{n}$ and $\bar{\alpha}_{n}:=\prod_{s=1}^{n}\alpha_{s}$ represents the noise level at each step. 
At the final time step $N$, $\boldsymbol{z}_{N}\sim\mathcal N(\boldsymbol{0},\boldsymbol{I})$ has a standard isotropic Gaussian distribution.
For model optimization, we employ the reweighted noise estimation training objective \cite{DDPM,DiffWave,rombach2022high}:
\begin{align}
\label{trainingobjective}
L_{n}(\theta)=\mathbb{E}_{\boldsymbol{z}_{0},\boldsymbol{\epsilon},n}\left \| \boldsymbol{\epsilon} - \boldsymbol{\epsilon}_{\theta}(\boldsymbol{z}_{n},n,\boldsymbol{E}^{x}) \right\|^2_{2},
\end{align}
where $\boldsymbol{E}^{x}$ is the embedding of the audio waveform $x$ produced by the pretrained audio encoder $f_{\text{audio}}(\cdot)$ in CLAP. In the reverse process, starting from Gaussian noise distribution $p(\boldsymbol{z}_{N})\sim\mathcal N(\boldsymbol{0},\boldsymbol{I})$ and the text embedding $\boldsymbol{E}^{y}$, a denoising process conditioned on $\boldsymbol{E}^{y}$ gradually generates the audio prior $\boldsymbol{z}_{0}$ by the following process:
\begin{align}
p_{\theta}(\boldsymbol{z}_{0:N}|\boldsymbol{E}^{y})&=p(\boldsymbol{z}_{N})\prod_{t=n}^{N}p_{\theta}(\boldsymbol{z}_{n-1}|\boldsymbol{z}_{n},\boldsymbol{E}^{y}) \\
\label{singlereversestep}
p_{\theta}(\boldsymbol{z}_{n-1}|\boldsymbol{z}_{n},\boldsymbol{E}^{y})&=\mathcal{N}(\boldsymbol{z}_{n-1};\boldsymbol{\mu}_{\theta}(\boldsymbol{z}_{n},n,\boldsymbol{E}^{y}),\boldsymbol{\sigma}_{n}^{2}\boldsymbol{I}).
\end{align}
The mean and variance are parameterized as~\cite{DDPM}: 
\begin{align}
\boldsymbol{\mu}_{\theta}(\boldsymbol{z}_{n},n,\boldsymbol{E}^{y})&=\frac{1}{\sqrt{\alpha_{n}}}(\boldsymbol{z}_{n}-\frac{\beta_{n}}{\sqrt{1-\bar{\alpha}_{n}}}\boldsymbol{\epsilon}_{\theta}(\boldsymbol{z}_{n},n,\boldsymbol{E}^{y})) \\
\boldsymbol{\sigma}_{n}^{2}&=\frac{1-\bar{\alpha}_{n-1}}{1-\bar{\alpha}_{n}}\beta_{n}
\end{align}
where $\boldsymbol{\epsilon}_{\theta}(\boldsymbol{z}_{n},n,\boldsymbol{E}^{y})$ is the predicted generation noise, and $\boldsymbol{\sigma}_{1}^{2}=\beta_{1}$. In the training stage, we learn the generation of an audio prior $\boldsymbol{z}_{0}$ given the cross-modal representation $\boldsymbol{E}^{x}$ of an audio sample $x$. Then, in TTA generation, we provide the text embeddings $\boldsymbol{E}^{y}$ to predict the noise $\boldsymbol{\epsilon}_{\theta}(\boldsymbol{z}_{n},n,\boldsymbol{E}^{y})$. Built on the CLAP embeddings, our LDM realizes TTA generation without text supervision in the training stage. We provide the details of network architecture in Appendix~\ref{app:LDMArchitecture}.

\subsection{Conditioning Augmentation}
\label{CA}

In text-to-image generation, diffusion-based models have demonstrated an ability to capture the fine-grained details between objects and backgrounds~\cite{DALLE2,Imagen,CompositionalDDPM}. One of the reasons for this success is the large-scale language-image training pairs, such as $400$ million image-text pairs in the LAION dataset~\cite{schuhmann2021laion}. For TTA generation, it is also desired to generate compositional audio signals whose relationships are consistent with natural language descriptions. However, the scale of available language-audio datasets is not comparable to that of language-image datasets. For data augmentation, AudioGen~\cite{kreuk2022audiogen} use a mixup strategy which mixes pairs of audio samples and concatenates their respective processed text captions to form new paired data. In our work, as shown in Equation~\ref{trainingobjective}, we provide the audio only embedding $\boldsymbol{E}^{x}$ as conditioning information when training LDMs, we can implement data augmentation on audio only signals instead of needing to augment language-audio pairs. Specifically, we perform mixup augmentation on audio $x_{1}$ and $x_{2}$ by:  
\begin{align}
\label{mixup}
x_{1,2}=\lambda x_{1}+(1-\lambda)x_{2},
\end{align}
where $\lambda$ is a scaling factor varying between $\left[0, 1\right]$ sampled from a Beta distribution $\mathcal B(5,5)$~\cite{gong2021psla}. Here we do not need to consider the corresponding text description $y_{1,2}$, since text information is not needed during LDM training. By mixing audio pairs, we increase the number of training data pairs $(\boldsymbol{z}_{0}, \boldsymbol{E}^{x})$ for LDMs, which makes LDMs robust to CLAP embeddings. In the sampling process, given the text embedding $\boldsymbol{E}^{y}$ from unseen language descriptions, LDMs are expected to generate the corresponding audio prior $\boldsymbol{z}_{0}$.

\subsection{Classifier-free Guidance}
\label{CFG}

\begin{figure}[htbp]
    \centering
    \includegraphics[width=1.0\linewidth]{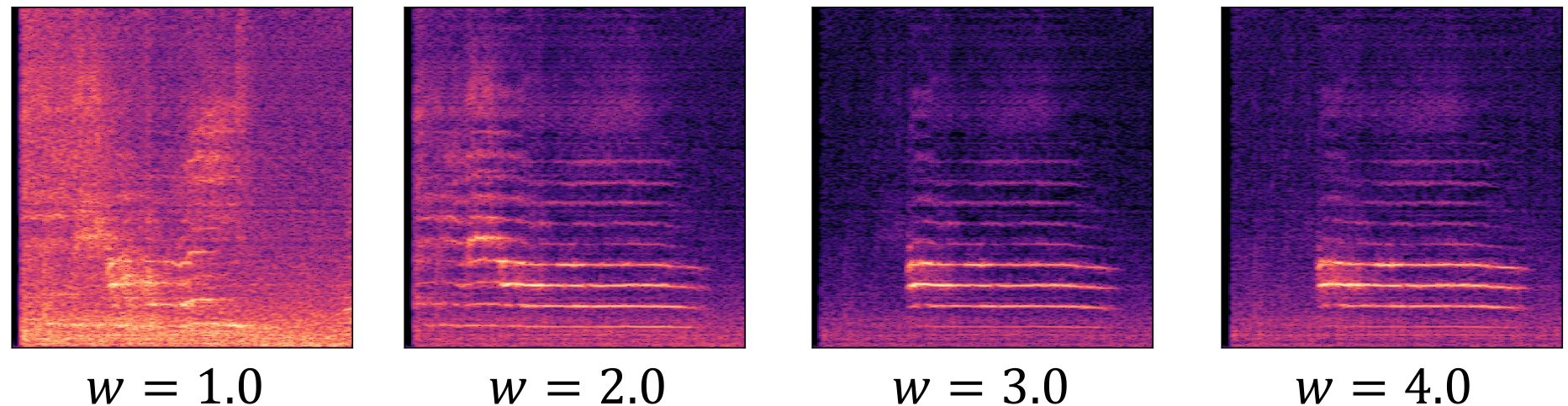}
    \caption{The samples generated with different scales of the classifier-free guidance. The text prompt is \textit{``A cat is meowing"}.}
    \label{fig:my_label}
\end{figure}

For diffusion models, controllable generation can be achieved by introducing guidance at each sampling step. After classifier guidance~\cite{SGM,ImprovedDDPM}, classifier-free guidance~\cite{CFG,Glide} (CFG) has been the state-of-the-art technique for guiding diffusion models. During training, we randomly discard our condition $\boldsymbol{E}^{x}$ with a fixed probability, e.g., $10\%$ to train both the conditional LDMs $\boldsymbol{\epsilon}_{\theta}(\boldsymbol{z}_{n},n,\boldsymbol{E}^{x})$ and the unconditional LDMs $\boldsymbol{\epsilon}_{\theta}(\boldsymbol{z}_{n},n)$. In generation, we use text embedding $\boldsymbol{E}^{y}$ as condition and perform sampling with a modified noise estimation $\hat{\boldsymbol{\epsilon}}_{\theta}(\boldsymbol{z}_{n},n,\boldsymbol{E}^{y})$:
\begin{align}
\label{mixup}
\hat{\boldsymbol{\epsilon}}_{\theta}(\boldsymbol{z}_{n},n,\boldsymbol{E}^{y}) = w\boldsymbol{\epsilon}_{\theta}(\boldsymbol{z}_{n},n)+(1-w)\boldsymbol{\epsilon}_{\theta}(\boldsymbol{z}_{n},n,\boldsymbol{E}^{y}),
\end{align}
where $w$ determines the guidance scale.
Compared with AudioGen~\cite{kreuk2022audiogen}, we have two differences. First, they leverage CFG on a transformer-based auto-regressive model, while our LDMs retain the theoretical formulation behind the CFG~\cite{CFG}. Second, our text embedding $\boldsymbol{E}^{y}$ is extracted from unprocessed natural language and therefore enables CFG to make use of the detailed text descriptions as guidance for audio generation. However, AudioGen removed the text details showing spatial or temporal relationships with text preprocessing methods. 

\subsection{Decoder}
\label{Decoder}

We use VAE to compress the mel-spectrogram $\boldsymbol{X}\in \mathbb{R}^{T\times F}$ into a small latent space $\boldsymbol{z}\in \mathbb{R}^{C\times \frac{T}{r}\times \frac{F}{r}}$, where $r$ is the compression level of the latent space. Our VAE is composed of an encoder and a decoder with stacked convolutional modules. In the training objective, we adopt a reconstruction loss, an adversarial loss, and a Gaussian constraint loss. We provide the detailed architecture and training methods in Appendix~\ref{app:VAE}. In the sampling process, the decoder is used to reconstruct the mel-spectrogram $\boldsymbol{\hat{X}}$ from the audio prior $\boldsymbol{\hat{z}}_{o}$ generated from LDMs. To explore a compression level $r$ that achieves a small latent space for LDMs without sacrificing sample quality, we test a group of values $r\in \left[1,2, 4,8,16\right]$, and take $r\vequalsignnospace4$ as our default setting because of its high computational efficiency and generation quality. Moreover, as we conduct conditioning augmentation for LDMs, we implement data augmentation with Equation~\ref{mixup} for VAE as well in order to guarantee the reconstruction quality of generated compositional samples. For vocoder, we employ HiFi-GAN~\cite{kong2020hifi} to generate the audio sample $\hat{x}$ from the reconstructed mel-spectrogram $\boldsymbol{\hat{X}}$. The training details are shown in Appendix~\ref{app:HiFi-GAN}.


\section{Text-Guided Audio Manipulation}
\label{AudioLDM}

\begin{figure}[tbp]
    \centering
    \includegraphics[width=\linewidth]{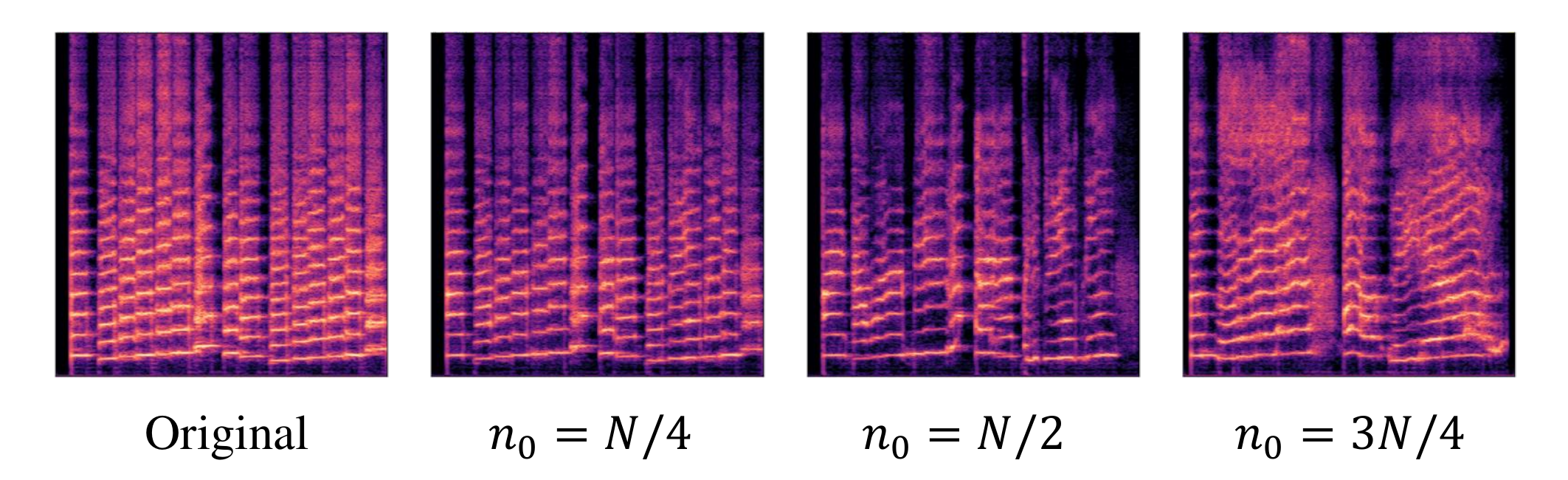}
    \caption{The manipulation result with different starting points $n_0$ of the shallow reverse process. The original signal is \textit{Trumpet}, and the text prompt for style transfer is \textit{Children Singing}.}
    \label{fig:style-transfer-demo}
\end{figure}

\textbf{Style Transfer}
\label{AST}
Given a source audio sample $x^{src}$, we can calculate its noisy latent representation $\boldsymbol{z}_{n_{0}}$ with a predefined time step $n_{0}\leq N$ according to the forward process shown in Equation~\ref{forwardprocess}. By utilizing $\boldsymbol{z}_{n_{0}}$ as the starting point of the reverse process of a pretrained AudioLDM model, we enable the manipulation of audio $x^{src}$ with text input $y$ with a shallow reverse process $p_{\theta}(\boldsymbol{z}_{0:n_{0}}|\boldsymbol{E}^{y})$:
\begin{align}
\label{shallowreverse}
p_{\theta}(\boldsymbol{z}_{0:n_{0}}|\boldsymbol{E}^{y})&=p(\boldsymbol{z}_{n_{0}})\prod_{n=1}^{n_{0}}p_{\theta}(\boldsymbol{z}_{n-1}|\boldsymbol{z}_{n},\boldsymbol{E}^{y}),
\end{align}
where $n_{0}$ controls the manipulation results. If we define a $n_{0}\approx N$, the information provided by source audio will not be retained and the manipulation would be similar to TTA generation. We show the effect of $n_{0}$ in Figure~\ref{fig:style-transfer-demo}, where larger manipulations can be seen in the setting of $n_{0}=3N/4$.

\textbf{Inpainting and Super-Resolution}
\label{AI}
Both audio inpainting and audio super-resolution refer to generating the missing part given the observed part $x^{ob}$. We explore these tasks by incorporating the observed part in latent representation $\boldsymbol{z}^{ob}$ into the generated latent representation $\boldsymbol{z}$. Specifically, in reverse process, starting from $p(\boldsymbol{z}_{N})\sim\mathcal N(\boldsymbol{0},\boldsymbol{I})$, after each inference step shown in Equation~\ref{singlereversestep}, we modify the generated $\boldsymbol{z}_{n-1}$ with:
\begin{align}
\label{restoration}
\boldsymbol{z}^{\prime}_{n-1}&=(1-\boldsymbol{m})\odot\boldsymbol{z}_{n-1}+\boldsymbol{m}\odot\boldsymbol{z}^{ob}_{n-1},
\end{align}
where $\boldsymbol{z}^{\prime}$ is the modified latent representation, $\boldsymbol{m}\in \mathbb{R}^{\frac{T}{r}\times \frac{F}{r}}$ denotes an observation mask in latent space, $\boldsymbol{z}^{ob}_{n-1}$ is obtained by adding noise on $\boldsymbol{z}^{ob}$ with the forward process shown by Equation~\ref{forwardprocess}. 

The values of observation mask $\boldsymbol{m}$ depend on the observed part of a mel-spectrogram $\boldsymbol{X}$. As we adopt a convolutional structure in VAE to learn the latent representation $\boldsymbol{z}$, we can roughly retain the spatial correspondency in mel-spectrogram, as it is shown in Figure~\ref{fig:demo-spatial-correspondancy} in Appendix~\ref{app:VAE}. Therefore, if a time-frequency bin $\boldsymbol{X}_{t,f}$ is observed, we set the observation mask $\boldsymbol{m}_{\frac{t}{r},\frac{f}{r}}$ in latent space as $1$.
By using $\boldsymbol{m}$ to denote the generation part and observation part in $\boldsymbol{z}$, according to Equation~\ref{restoration}, we can generate the missing information conditioned on the text prompt with TTA models, while retaining the ground-truth observation $\boldsymbol{z}^{ob}$.


\begin{table*}[tbp]
\centering
\scriptsize
\begin{tabular}{ccccc|cccc|cc}
\toprule
    Model    & Text Data & Use CLAP & Params & Duration~(h) & FD~$\downarrow$  & IS~$\uparrow$   & KL~$\downarrow$ & FAD~$\downarrow$ & OVL~$\uparrow$ & REL~$\uparrow$   \\
\midrule
Ground truth & - & - & - & - & - & - & - & - & $83.61_{\pm 1.1}$ & $80.11_{\pm 1.2}$ \\
DiffSound$^{\dagger}$~\cite{yang2022diffsound}   & \cmark & \xmark           & $400$M  & $5420$ & $47.68$ & $4.01$ & $2.52$ & $7.75$ & $45.00_{\pm 2.6}$ & $43.83_{\pm 2.3}$ \\
AudioGen$^{\dagger}$~\cite{kreuk2022audiogen}      & \cmark & \xmark &$285$M & $8067$  & -    &  -    & $2.09$  & $3.13$ & - & - \\
\midrule
AudioLDM-S-Full-RoBERTa   & \cmark   & 
\xmark      & $181$M & $145$  & $32.13$  & $4.02$ & $3.25$ & $5.89$ & - & - \\
AudioLDM-S    & \xmark  & 
\cmark       & $181$M & $145$  & $29.48$  & $6.90$ & $1.97$ & $2.43$ & $63.41_{\pm 1.4}$ & $64.83_{\pm 0.9}$ \\
AudioLDM-L    & \xmark   & 
\cmark     & $739$M & $145$ & $27.12$  & $7.51$ & $1.86$ & $2.08$ 
 & $64.30_{\pm 1.6}$ & $64.72_{\pm 1.6}$\\
AudioLDM-S-Full & \xmark & 
\cmark   & $181$M & $8886$ & $23.47$  & $7.57$ & $1.98$ & $2.32$ & - & - \\
AudioLDM-L-Full & \xmark & 
\cmark  & $739$M & $8886$ & $\mathbf{23.31}$  & $\mathbf{8.13}$ & $\mathbf{1.59}$ & $\mathbf{1.96}$ & $\mathbf{65.91}_{\pm 1.0}$ & $\mathbf{65.97}_{\pm 1.6}$ \\
\bottomrule
\end{tabular}
\caption{The comparison between AudioLDM and baseline TTA generation models. Evaluation is conducted on AudioCaps test set. The symbol $^{\dagger}$ marks industry-level computation. DiffSound is trained on $32$ V$100$ GPUs and AudioGen is trained on $64$ A$100$ GPUs, while AudioLDM models are trained on a single GPU, RTX $3090$ or A$100$. The AS and AC stand for AudioSet and AudioCaps datasets respectively. The results of AudioGen are employed from ~\citep{kreuk2022audiogen} since their implementation has been not publicly available.}
\label{tab: AudioCapResults}
\end{table*}

\section{Experiments}
\label{Experiments}

\textbf{Training dataset} The datasets we used in this paper includes AudioSet~(AS)~\cite{gemmeke2017audio}, AudioCaps~(AC)~\cite{kim2019audiocaps}, Freesound~(FS)\footnote{\url{https://freesound.org/}}, and BBC Sound Effect library~(SFX)\footnote{\url{https://sound-effects.bbcrewind.co.uk/search}}. AS is currently the largest audio dataset, with $527$ labels and over $5,000$ hours of audio data. AC is a much smaller dataset with around $49,000$ audio clips and text descriptions. Most of the data in AudioSet and AudioCaps are in-the-wild audio from YouTube, so the quality of the audio is not guaranteed. To expand the dataset, especially with high-quality audio data, we crawl the data from the FreeSound and BBC SFX datasets, which have a wide range of categories such as music, speech, and sound effects. We show our detailed data processing methods and training configuration in Appendix~\ref{app:TrainingDetails}.

\textbf{Evaluation dataset} We evaluate the model on both AC and AS. Each audio clip in AC has $5$ text captions. We generate the evaluation set by randomly selecting one of them as text condition. Because the authors of AC intentionally remove the audio with the label related to music~\cite{kim2019audiocaps}, to evaluate model performance with a wider range of sound, we randomly select $10\%$ audio samples from AS as another evaluation set. Since AS does not contain text descriptions, we use the concatenation of labels as text descriptions, such as \textit{Speech, hip hop music, and crowd cheering}. 



\textbf{Evaluation methods} We perform both objective evaluation and human subjective evaluation. The main metrics we use for objective evaluation include \textbf{frechet distance~(FD)}, \textbf{inception score~(IS)}, and \textbf{kullback–leibler~(KL) divergence}. Similar to the frechet inception distance in image generation, the FD in audio indicates the similarity between generated samples and target samples. IS is effective in evaluating both sample quality and diversity. KL is measured at a paired sample level and averaged as the final result. All of these three metrics are built upon a state-of-the-art audio classifier PANNs~\cite{kong2020panns}. To compare with~\cite{kreuk2022audiogen}, we also adopt the frechet audio distance~(FAD)~\cite{kilgour2019frechet}. FAD has a similar idea to FD but it uses VGGish~\cite{vggish_hershey2017cnn} as a classifier which may have inferior performance than PANNs. To better measure the generation quality, we choose FD as the main evaluation metric in this paper. For subjective evaluation, we recruit six audio professionals to carry on a rating process following~\cite{kreuk2022audiogen,yang2022diffsound}. Specifically, the generated samples are rated based on i) \textbf{overall quality~(OVL)}; and ii) \textbf{relevance to the input text~(REL)} between a scale of $1$ to $100$. We include the details of human evaluation in Appendix~\ref{app:TrainingDetails}. We open-source our evaluation pipeline to facilitate reproducibility\footnote{\url{https://github.com/haoheliu/audioldm_eval}}.

\textbf{Models} We employ two recently proposed TTA systems, DiffSound~\cite{yang2022diffsound} and AudioGen~\cite{kreuk2022audiogen} as our baseline models. DiffSound is trained on AS and AC datasets with around $400$M parameters. AudioGen is trained on AS, AC, and eight other datasets with around $285$M parameters. Since AudioGen has not released publicly available implementation, we reuse the KL and FAD results reported in their paper. We train two AudioLDM models. One is a small model named AudioLDM-S, which has $181$M parameters, and the other is a large model named AudioLDM-L with $739$M parameters. We describe the details of UNet architecture in Appendix~\ref{app:LDMArchitecture}. To demonstrate the advantage of our method, we simply train these two models only with the AC dataset. Moreover, to explore the effect of the scale of training data, we develop an AudioLDM-L-Full model which is trained on AC, AS, FreeSound, and BBC SFX datasets.


\subsection{Results}

We show the main evaluation results on the AC test set in Table~\ref{tab: AudioCapResults}. Given the single training dataset AC, AudioLDM-S can achieve better generation results than the baseline models on both objective and subjective evaluations, even with smaller model size. By expanding model capacity with AudioLDM-L, we further improve the overall results. Then, by incorporating AS and the two other datasets into training, our model AudioLDM-L-Full achieves the best quality, with an FD of $23.31$. 
Although RoBERTa and CLAP have the same text encoder structure, CLAP has an advantage in that it decouples audio-text relationship learning from generative model training. This decoupling is intuitive as CLAP has already modelled the relationship between audio and text by aligning their embedding spaces. On the other hand, AudioLDM-S-Full-RoBERTa, in which the text encoder only represents textual information, requires the model to learn the text-audio relationships while simultaneously learning the audio generation process. Additionally, our CLAP-based method allows for model training using audio-only data. Therefore, using Roberta without pretraining with CLAP may increase the difficulty of training.

\begin{figure*}[tbp]
    \centering
    \includegraphics[width=1.0\linewidth]{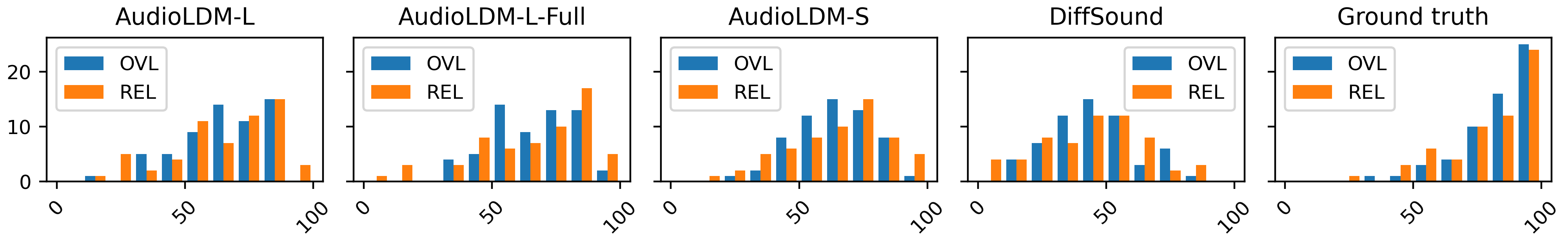}
    \caption{The histogram of the human evaluation result. The horizontal axis and vertical axis represent the rating score and frequency, respectively. \textit{OVL} denotes the overall quality of audio files and \textit{REL} denotes the relation between text and generated audio. Both OVL and REL are scored on a scale of $1$ to $100$. Scores on each evaluation file are averaged among all the raters.}
    \label{fig:hist-human-evaluation}
\end{figure*}

\begin{figure}[tbp]
    \centering
    \includegraphics[width=1.0\linewidth]{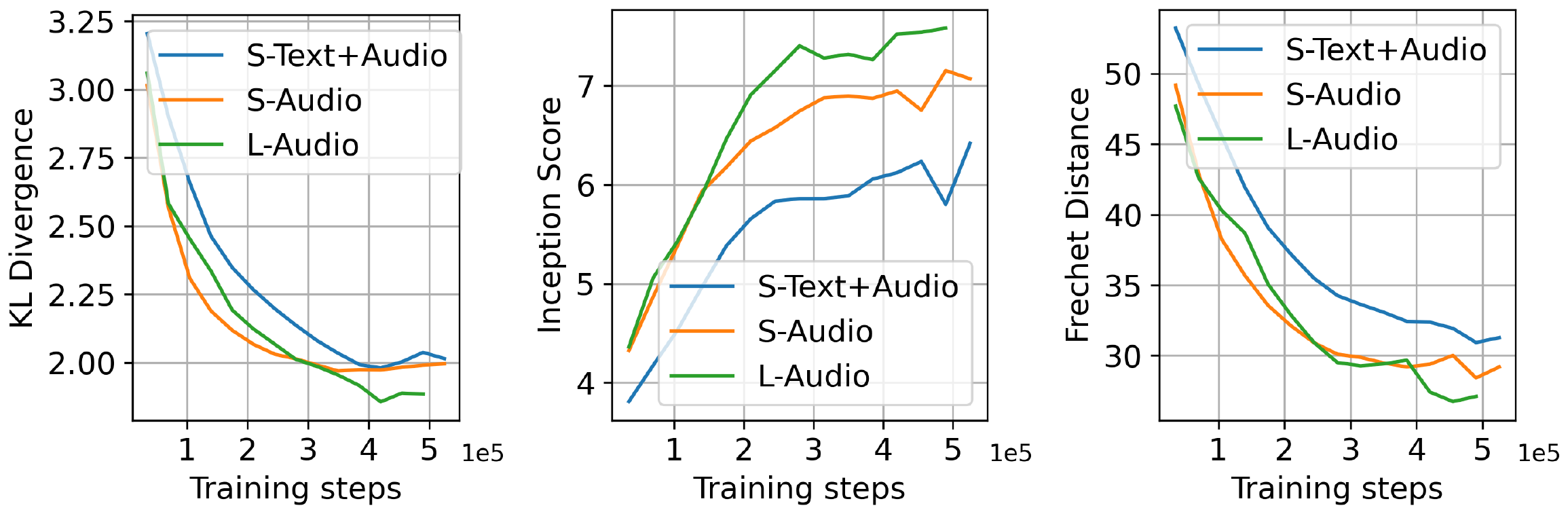}
    \caption{The comparison of various evaluation metrics evaluated in the training process of i) AudioLDM-S trained with text embedding (S-Text+Audio) ii) AudioLDM-S (S-Audio), and iii) AudioLDM-L (L-Audio).}
    \label{fig:training-steps-fd-is-kl}
\end{figure}

Our human evaluation shows a similar trend as other evaluation metrics. Our proposed methods have OVL and REL of around $64$, outperforming DiffSound with OVL of $45.00$ and REL of $43.83$ by a large margin. On the AudioLDM model size, we notice that the larger model is advantageous for the overall audio qualities. After scaling up the training data, both OVL and REL show significant improvements.
Figure~\ref{fig:hist-human-evaluation} shows the score statistic of different models averaged between all the raters. We notice our model is more concentrated on the higher scores compared with DiffSound. Our spam cases, which are randomly selected real recordings, show high scores, indicating the rating result is reliable.


To perform the evaluation on audio data that could include music, we further evaluate our model on the AS evaluation set. We compare our method with DiffSound and show the results in Table~\ref{tab: AudioSetResults}. Our three AudioLDM models show a similar trend as they perform on the AC test set. We can outperform the DiffSound baseline by a large margin on all the metrics. 

\begin{table}[htbp]
\small
\centering
\begin{tabular}{ccccc}
\toprule
       Model            & FD~$\downarrow$   & IS~$\uparrow$   & KL~$\downarrow$  \\
\midrule
DiffSound          & $50.40$ & $4.19$  & $3.63$ \\
AudioLDM-S      & $28.08$ & $6.78$ & $2.51$  \\
AudioLDM-L      & $27.51$ & $7.18$ & $2.49$ \\
AudioLDM-L-Full & $\mathbf{24.26}$ & $\mathbf{7.67}$ & $\mathbf{2.07}$ \\
\bottomrule
\end{tabular}
\caption{The evaluation results on the AudioSet evaluation set.}
\label{tab: AudioSetResults}
\end{table}

\textbf{Conditioning Information}
As we train LDMs conditioned on the audio embedding $\boldsymbol{E}^{x}$ but provide the text embedding $\boldsymbol{E}^{y}$ to LDMs in TTA generation, a natural concern is that if stronger results could be achieved by directly using the text embedding as training condition. We conduct experiments and show the results in Table~\ref{tab: conditioning-text-audio}. For a fair comparison, we also conduct data augmentation and we adopt the strategy from AudioGen. Specifically, we use the same mixing method for audio pairs shown in Section~\ref{CA}, and concatenate two text captions as conditioning information. Table~\ref{tab: conditioning-text-audio} shows by training LDMs on $\boldsymbol{E}^{x}$, we can achieve better results than training with $\boldsymbol{E}^{y}$. 

\begin{table}[htbp]
\centering
\footnotesize
\begin{tabular}{cccccc}
\toprule
    Model     & Text & Audio  & FD~$\downarrow$   & IS~$\uparrow$   & KL~$\downarrow$  \\
\midrule
AudioLDM-S & \cmark & \cmark & $31.26$ & $6.35$ & $2.01$ \\
AudioLDM-S & \xmark & \cmark & $\mathbf{29.48}$  & $\mathbf{6.90}$ & $\mathbf{1.97}$ \\
AudioLDM-S-Full & \cmark & \cmark & $27.20$ & $7.52$ & $2.38$ \\
AudioLDM-S-Full & \xmark & \cmark & $\mathbf{23.47}$ & $\mathbf{7.57}$ & $\mathbf{1.98}$ \\
AudioLDM-L-Full & \cmark & \cmark & $25.79$ & $7.95$ & $2.26$ \\
AudioLDM-L-Full & \xmark & \cmark & $\mathbf{23.31}$ & $\mathbf{8.13}$ & $\mathbf{1.59}$ \\
\bottomrule
\end{tabular}
\caption{The comparison between text embedding and audio embedding as conditioning information on the training of LDMs.}
\label{tab: conditioning-text-audio}
\end{table}

We believe the primary reason for the result in Table~\ref{tab: conditioning-text-audio} is that text embedding cannot represent the generation target as good as audio embedding. 
Firstly, due to the ambiguity and complexity of sound, the text caption is difficult to be accurate and comprehensive. Different human annotators may have different perceptions and descriptions over the same audio, which make training with text-audio pair less stable than with audio only.
Moreover, some of the captions are at a highly-abstracted level and cannot correctly describe the audio content. For example, there is an audio in the BBC SFX dataset with caption \textit{Boats: Battleships-5.25 conveyor space}, which is even difficult for humans to imagine how it sounds. This quality of language-audio pairs may hinder model optimization. 
By comparison, if we use $\boldsymbol{E}^{x}$ from CLAP latents as a condition, it is extracted directly from the audio signal and is aligned with ideally the best text caption, which enables us to provide strong conditioning information to LDMs without considering the noisy labeled text description.
Figure~\ref{fig:training-steps-fd-is-kl} shows sample quality as a function of training progress. We notice that i) training with audio embedding can lead to significantly better results than text embedding throughout the entire training process; and ii) larger models may converge more slowly but can achieve better final performance.


\textbf{Compression Level} We study the effect of compression level $r$ on generation quality. Table~\ref{tab: compressionratio} shows the performance comparison with $r\vequalsignnospace4,8,16$. We observe a decreasing trend with the increase of compression levels. Nevertheless, in the setting of $r\vequalsignnospace16$ where we compress the $64$-band mel-spectrogram into only $4$ dimensions in the frequency axis, our performance is still on par with AudioGen on KL, and better than DiffSound on all the metrics.

\begin{table}[htbp]
\centering
\small
\begin{tabular}{ccccc}
\toprule
Model         & $r$ & FD~$\downarrow$   & IS~$\uparrow$   & KL~$\downarrow$   \\
\midrule
AudioLDM-S & $4$          & $\mathbf{29.48}$ & $\mathbf{6.90}$ & $\mathbf{1.97}$ \\
AudioLDM-S & $8$          & $33.50$  & $6.13$ & $2.04$ \\
AudioLDM-S & $16$         & $34.32$  & $5.68$ & $2.09$ \\
\bottomrule
\end{tabular}
\caption{The effect of the compression level on AudioLDM.}
\label{tab: compressionratio}
\end{table}

If we set the compression level as $r\vequalsignnospace1$, which means we directly generate mel-spectrogram from CLAP embeddings, the training process is difficult to implement on a single RTX $3090$ GPU. Similar results happen on $r\vequalsignnospace2$. Moreover, the inference speed will be low with $r\vequalsignnospace1,2$. In our studies, $r\vequalsignnospace4$ achieves high generation quality while reducing the computational load to a reasonable level. Hence, we use it as the default setting in our experiments.

\textbf{Text-Guided Audio Manipulation} We show the performance of our text-guided audio manipulation methods on two tasks: super-resolution and inpainting. Specifically, for super-resolution, we upsample the audio signal from $8$~kHz to $16$~kHz sampling rate. For the inpainting task, we remove the audio signal between $2.5$ and $7.5$ seconds and refill this part by inpainting. Since most studies on audio super-resolution work on speech signal~\cite{liu2021voicefixer, liu2022neural}, we demonstrate our results on both AudioCaps, and a speech dataset VCTK~\cite{vctk-yamagishi2019cstr}, which is a multi-speaker speech dataset. For super-resolution, we employ two models AudioUNet~\cite{kuleshov2017audio} and NVSR~\cite{liu2022neural} as baseline models, and employ log-spectral distance (LSD)~\cite{heming-towards-sr-wang2021towards} as the evaluation metric for comparison. For the inpainting task, we use FAD as a metric and establish a baseline for this task. 
\begin{table}[tbp]
\centering
\small
\begin{tabular}{cccc}
\toprule
Task          & \multicolumn{2}{c}{Super-resolution}  &  Inpainting     \\
\midrule
Dataset       & AudioCaps            & VCTK                 & AudioCaps            \\
\midrule
Unprocessed   &          $2.76$            &      $2.15$                &   $10.86$                   \\
\citet{kuleshov2017audio}     & -                    & $1.32$                 & -                    \\
\citet{liu2022neural}          & -                    & $\mathbf{0.78}$                 & -                    \\
AudioLDM-S &         $1.59$             &     $1.12$                 &   $2.33$                   \\
AudioLDM-L & $\mathbf{1.43}$ & $0.98$ & $\mathbf{1.92}$ \\
\bottomrule
\end{tabular}
\caption{Performance comparisons on zero-shot super-resolution and inpainting, which are evaluated by LSD and FAD, respectively.}
\label{tab: audiomanipulation}
\end{table}
Table~\ref{tab: audiomanipulation} shows that AudioLDM can outperform the strong AudioUNet baseline, but the result is not as good as NVSR~\cite{liu2022neural}. 
Recall that AudioLDM is a model trained on a diverse set of audio signals, including those with heavy background noise. This can lead to the presence of white noise or other non-speech sound events in the output of our super-resolution process, potentially reducing performance. 
Nevertheless, our contribution could open the door to achieving text-guided audio manipulation with the TTA system in a zero-shot way. Further improvements could be expected based on our benchmark results. We provide several samples of our results in Appendix~\ref{app:demos}. 

\subsection{Ablation Study}

Table~\ref{tab: ablationstudy} shows the result of our ablation study on AudioLDM-S. By simplifying the attention mechanism in UNet into a one-layer multi-head self-attention~(\textit{w.~Simple attn}), the performance in each metric will have a notable decrease, which indicates complex attention mechanism is preferred. Also, we notice the widely used balanced sampling strategy~\cite{gong2021psla, liu2022ontology} in audio classification does not show improvement in TTA~(\textit{w.~Balance samp}). Conditional augmentation~(see Section~\ref{CA}) shows improvement in the subjective evaluation, but it does not show improvement in the objective evaluation metrics~(\textit{w.~Cond aug}). 
The reason could be that conditioning augmentation generates training data that is not representative of the AudioCaps dataset, resulting in model outputs that are not well-aligned with the evaluation data, ultimately leading to lower metric scores. Nevertheless, conditioning augmentation can improve two subjective metrics and we still recommend using it as a data augmentation technique.

\begin{table}[htbp]
\small
\centering
\begin{tabular}{lccc|cc}
\toprule
       Setting            & FD$\downarrow$   & IS$\uparrow$   & KL$\downarrow$  & OVL~$\uparrow$ & REL~$\uparrow$ \\
\midrule
AudioLDM-S      & $\mathbf{29.48}$ & $\mathbf{6.90}$ & $\mathbf{1.97}$ & $63.41$ & $64.83$ \\
\textit{w.~Simple attn}             & $33.12$ & $6.15$ & $2.09$ & - & - \\
\textit{w.~Balance samp} &  $34.05$     &   $6.21$   &  $2.16$   & - & - \\
\textit{w.~Cond aug}           & $31.88$ & $6.25$ & $2.02$ & $\mathbf{64.49}$ & $\mathbf{65.01}$  \\
\bottomrule
\end{tabular}
\caption{The ablation study on the attention mechanism, the balance sampling technique for training data, and the conditioning augmentation algorithm.}
\label{tab: ablationstudy}
\end{table}

\textbf{DDIM Sampling Step} The number of inference steps in the reverse process of DDPMs can directly affect the generation quality \cite{DDPM,SGM}. Generally, the sample quality can be improved with an increase in the number of sampling steps and computational load at the same time. We explore the effect of the DDIM~\cite{song2020denoising} sampling steps on our latent diffusion model. Table~\ref{tab: DDIMsampling} shows that more sampling steps lead to better quality. With enough sampling steps such as $100$, the gain of adding sampling steps becomes less significant. The result of $200$ steps is only slightly better than that of $100$ steps.

\begin{figure}[tbp]
    \centering
    \includegraphics[width=0.9\linewidth]{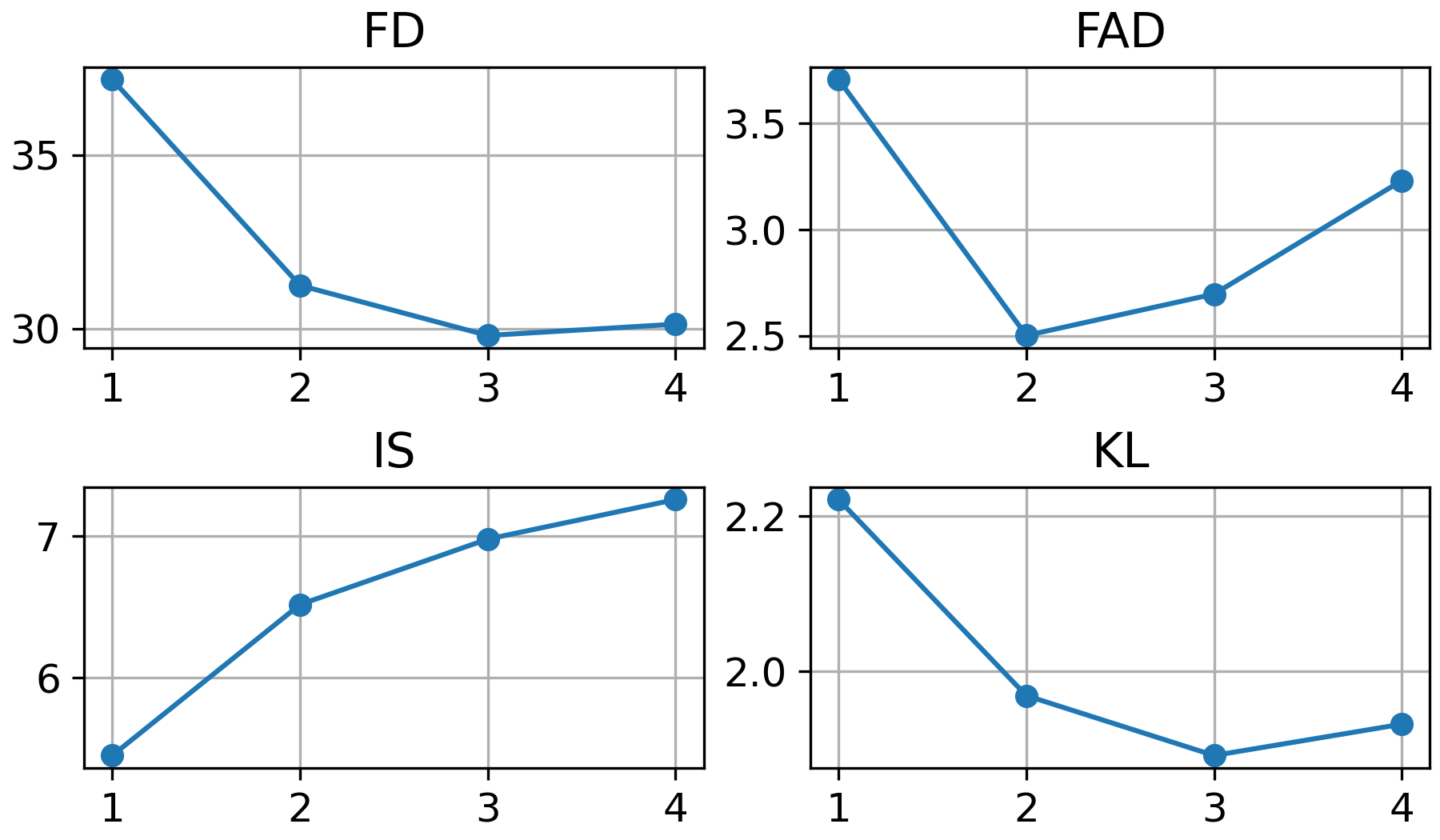}
    \caption{The comparison between different classifier-free guidance scales (on the horizontal axis) for the AudioLDM-S model trained on AudioCaps.}
    \label{fig:classifier-free-guidance}
    \vspace{-2mm}
\end{figure}

\begin{table}[tbp]
\centering
\small
\begin{tabular}{cccccc}
\toprule
  DDIM steps    & $10$    & $25$      & $50$     & $100$     & $200$   \\
\midrule
FD    &   $55.84$    &    $42.84$     &    $35.71$    &    $30.17$     &   $\mathbf{29.48}$    \\
IS    &   $4.21$    &    $5.91$     &   $6.51$     &    $6.85$     & 
 $\mathbf{6.90} $      \\ 
KL    &   $2.47$    &    $2.12$     &   $2.01$     &    $\mathbf{1.94}$     & $1.97$      \\
\bottomrule
\end{tabular}
\caption{Effect of sampling steps of LDMs with a DDIM sampler.}
\label{tab: DDIMsampling}
\vspace{-6mm}
\end{table}

\textbf{Guidance Scale} 
represents a trade-off between conditional generation quality and sample diversity. A suitable guidance scale can improve the consistency between generated samples and conditioning information at an acceptable cost of generation diversity. We show the effect of guidance scale $w$ on TTA in Figure~\ref{fig:classifier-free-guidance}. When $w=3$, we achieve the best results in both FD and KL, but not in FAD. We suppose the reason is the audio classifier in FAD is not as good as FD, as mentioned in Section~\ref{Experiments}. In this case, the improvement in the adherence to detailed language description may become misleading information to the classifier in FAD. Considering previous studies report FAD results instead of FD, we set $w=2$ for comparison, but also provide detailed effects of $w$ on FAD, FD, IS, and KL, respectively. 

\textbf{Case Study}~We conduct case study and show the generated results in Appendix~\ref{app:demos}, including style transfer~(see Figure~\ref{fig:demo-style-transfer-1}-\ref{fig:demo-style-transfer-3}), super-resolution~(see Figure~\ref{fig:demo-super-resolution}), inpainting~(see Figure~\ref{fig:demo-inpainting}-\ref{fig:demo-inpainting-prompt}), and text-to-audio generation~(see Figure~\ref{fig:demo-control-speech}-\ref{fig:demo-audioset-music}). Specifically, for text-to-audio, we demonstrate the controllability of AudioLDM, including the control of the acoustic environment, material, sound event, pitch, musical genres, and temporal orders.

\section{Conclusions}
\label{Conclusion}
We have presented a new method AudioLDM for text-to-audio~(TTA) generation, with contrastive language-audio pretraining~(CLAP) models and latent diffusion models~(LDMs). Our method is advantageous in generation quality, computational efficiency, and audio manipulations. With a single training dataset AudioCaps and a single GPU, AudioLDM achieves SOTA generation quality evaluated by both subjective and objective metrics. Moreover, AudioLDM enables zero-shot text-guided audio style transfer, super-resolution, and inpainting.

\section{Acknowledgement}
\label{sec:ack}
We would like to thank James King and Jinhua Liang for the useful discussion on the latent diffusion model. This research was partly supported by the British Broadcasting Corporation Research and Development~(BBC R\&D), Engineering and Physical Sciences Research Council (EPSRC) Grant EP/T019751/1 ``AI for Sound'', and a PhD scholarship from the Centre for Vision, Speech and Signal Processing (CVSSP), Faculty of Engineering and Physical Science (FEPS), University of Surrey. For the purpose of open access, the authors have applied a Creative Commons Attribution (CC BY) license to any Author Accepted Manuscript version arising.

\bibliography{maincontent}

\begin{thebibliography}{63}
\providecommand{\natexlab}[1]{#1}
\providecommand{\url}[1]{\texttt{#1}}
\expandafter\ifx\csname urlstyle\endcsname\relax
  \providecommand{\doi}[1]{doi: #1}\else
  \providecommand{\doi}{doi: \begingroup \urlstyle{rm}\Url}\fi

\bibitem[Andresen(1979)]{andresen1979new}
Andresen, U.
\newblock A new way in sound synthesis.
\newblock In \emph{Audio Engineering Society}. Audio Engineering Society, 1979.

\bibitem[Chen et~al.(2022{\natexlab{a}})Chen, Du, Zhu, Ma, Berg{-}Kirkpatrick,
  and Dubnov]{HTSAT}
Chen, K., Du, X., Zhu, B., Ma, Z., Berg{-}Kirkpatrick, T., and Dubnov, S.
\newblock {HTS-AT:} {A} hierarchical token-semantic audio transformer for sound
  classification and detection.
\newblock In \emph{{IEEE International Conference on Acoustics, Speech and
  Signal Processing}}, 2022{\natexlab{a}}.

\bibitem[Chen et~al.(2021)Chen, Zhang, Zen, Weiss, Norouzi, and Chan]{WaveGrad}
Chen, N., Zhang, Y., Zen, H., Weiss, R., Norouzi, M., and Chan, W.
\newblock Wavegrad: Estimating gradients for waveform generation.
\newblock In \emph{{International Conference on Learning Representations}},
  2021.

\bibitem[Chen et~al.(2022{\natexlab{b}})Chen, Tan, Wang, Pan, Mandic, He, and
  Zhao]{InferGrad}
Chen, Z., Tan, X., Wang, K., Pan, S., Mandic, D., He, L., and Zhao, S.
\newblock Infergrad: Improving diffusion models for vocoder by considering
  inference in training.
\newblock In \emph{{IEEE International Conference on Acoustics, Speech and
  Signal Processing}}, 2022{\natexlab{b}}.

\bibitem[Chen et~al.(2022{\natexlab{c}})Chen, Wu, Leng, Chen, Liu, Tan, Cui,
  Wang, He, Zhao, Bian, and Mandic]{ResGrad}
Chen, Z., Wu, Y., Leng, Y., Chen, J., Liu, H., Tan, X., Cui, Y., Wang, K., He,
  L., Zhao, S., Bian, J., and Mandic, D.
\newblock Resgrad: Residual denoising diffusion probabilistic models for text
  to speech.
\newblock \emph{arXiv preprint:2212.14518}, 2022{\natexlab{c}}.

\bibitem[Dhariwal \& Nichol(2021)Dhariwal and Nichol]{DiffusionBeatsGANs}
Dhariwal, P. and Nichol, A.
\newblock Diffusion models beat gans on image synthesis.
\newblock In \emph{{Conference on Neural Information Processing Systems}},
  2021.

\bibitem[Drossos et~al.(2020)Drossos, Lipping, and Virtanen]{Clotho}
Drossos, K., Lipping, S., and Virtanen, T.
\newblock Clotho: an audio captioning dataset.
\newblock In \emph{{IEEE International Conference on Acoustics, Speech and
  Signal Processing}}, 2020.

\bibitem[Engel et~al.(2020)Engel, Hantrakul, Gu, and Roberts]{engel2020ddsp}
Engel, J., Hantrakul, L., Gu, C., and Roberts, A.
\newblock Ddsp: Differentiable digital signal processing.
\newblock \emph{arXiv preprint:2001.04643}, 2020.

\bibitem[Gemmeke et~al.(2017)Gemmeke, Ellis, Freedman, Jansen, Lawrence, Moore,
  Plakal, and Ritter]{gemmeke2017audio}
Gemmeke, J.~F., Ellis, D.~P., Freedman, D., Jansen, A., Lawrence, W., Moore,
  R.~C., Plakal, M., and Ritter, M.
\newblock {AudioSet}: An ontology and human-labeled dataset for audio events.
\newblock In \emph{IEEE International Conference on Acoustics, Speech and
  Signal Processing}, pp.\  776--780. IEEE, 2017.

\bibitem[Gong et~al.(2021)Gong, Chung, and Glass]{gong2021psla}
Gong, Y., Chung, Y.-A., and Glass, J.
\newblock {PSLA}: Improving audio tagging with pretraining, sampling, labeling,
  and aggregation.
\newblock \emph{IEEE/ACM Transactions on Audio, Speech, and Language
  Processing}, 29:\penalty0 3292--3306, 2021.

\bibitem[Hershey et~al.(2017)Hershey, Chaudhuri, Ellis, Gemmeke, Jansen, Moore,
  Plakal, Platt, Saurous, Seybold, et~al.]{vggish_hershey2017cnn}
Hershey, S., Chaudhuri, S., Ellis, D.~P., Gemmeke, J.~F., Jansen, A., Moore,
  R.~C., Plakal, M., Platt, D., Saurous, R.~A., Seybold, B., et~al.
\newblock {CNN} architectures for large-scale audio classification.
\newblock In \emph{2017 IEEE International Conference on Acoustics, Speech and
  Signal Processing}, pp.\  131--135. IEEE, 2017.

\bibitem[Ho \& Salimans(2021)Ho and Salimans]{CFG}
Ho, J. and Salimans, T.
\newblock Classifier-free diffusion guidance.
\newblock In \emph{{NeurIPS Workshop on Deep Generative Models and Downstream
  Applications}}, 2021.

\bibitem[Ho et~al.(2020)Ho, Jain, and Abbeel]{DDPM}
Ho, J., Jain, A., and Abbeel, P.
\newblock Denoising diffusion probabilistic models.
\newblock In \emph{{Conference on Neural Information Processing Systems}},
  2020.

\bibitem[Ho et~al.(2022)Ho, Chan, Saharia, Whang, Gao, Gritsenko, Kingma,
  Poole, Norouzi, Fleet, and Salimans]{ImagenVideo}
Ho, J., Chan, W., Saharia, C., Whang, J., Gao, R., Gritsenko, A., Kingma,
  D.~P., Poole, B., Norouzi, M., Fleet, D.~J., and Salimans, T.
\newblock Imagen video: High definition video generation with diffusion models.
\newblock \emph{arXiv preprint:2210.02303}, 2022.

\bibitem[Isola et~al.(2017)Isola, Zhu, Zhou, and Efros]{isola2017image}
Isola, P., Zhu, J.-Y., Zhou, T., and Efros, A.~A.
\newblock Image-to-image translation with conditional adversarial networks.
\newblock In \emph{Proceedings of the IEEE/CVF Conference on Computer Vision
  and Pattern Recognition}, pp.\  1125--1134, 2017.

\bibitem[Karplus \& Strong(1983)Karplus and Strong]{karplus1983digital}
Karplus, K. and Strong, A.
\newblock Digital synthesis of plucked-string and drum timbres.
\newblock \emph{Computer Music Journal}, 7\penalty0 (2):\penalty0 43--55, 1983.

\bibitem[Kilgour et~al.(2019)Kilgour, Zuluaga, Roblek, and
  Sharifi]{kilgour2019frechet}
Kilgour, K., Zuluaga, M., Roblek, D., and Sharifi, M.
\newblock Fr{\'e}chet audio distance: A reference-free metric for evaluating
  music enhancement algorithms.
\newblock In \emph{INTERSPEECH}, pp.\  2350--2354, 2019.

\bibitem[Kim et~al.(2019)Kim, Kim, Lee, and Kim]{kim2019audiocaps}
Kim, C.~D., Kim, B., Lee, H., and Kim, G.
\newblock Audiocaps: Generating captions for audios in the wild.
\newblock In \emph{Proceedings of the 2019 Conference of the North American
  Chapter of the Association for Computational Linguistics: Human Language
  Technologies}, pp.\  119--132, 2019.

\bibitem[Kingma \& Ba(2014)Kingma and Ba]{kingma2014adam}
Kingma, D.~P. and Ba, J.
\newblock Adam: A method for stochastic optimization.
\newblock \emph{arXiv preprint:1412.6980}, 2014.

\bibitem[Kingma \& Welling(2013)Kingma and Welling]{kingma2013auto}
Kingma, D.~P. and Welling, M.
\newblock Auto-encoding variational bayes.
\newblock \emph{arXiv preprint:1312.6114}, 2013.

\bibitem[Kong et~al.(2020{\natexlab{a}})Kong, Kim, and Bae]{kong2020hifi}
Kong, J., Kim, J., and Bae, J.
\newblock Hifi-gan: Generative adversarial networks for efficient and high
  fidelity speech synthesis.
\newblock \emph{Advances in Neural Information Processing Systems},
  33:\penalty0 17022--17033, 2020{\natexlab{a}}.

\bibitem[Kong et~al.(2020{\natexlab{b}})Kong, Cao, Iqbal, Wang, Wang, and
  Plumbley]{kong2020panns}
Kong, Q., Cao, Y., Iqbal, T., Wang, Y., Wang, W., and Plumbley, M.~D.
\newblock {PANNs}: Large-scale pretrained audio neural networks for audio
  pattern recognition.
\newblock \emph{IEEE/ACM Transactions on Audio, Speech, and Language
  Processing}, 28:\penalty0 2880--2894, 2020{\natexlab{b}}.

\bibitem[Kong et~al.(2021{\natexlab{a}})Kong, Cao, Liu, Choi, and
  Wang]{kong2021decoupling}
Kong, Q., Cao, Y., Liu, H., Choi, K., and Wang, Y.
\newblock Decoupling magnitude and phase estimation with deep resunet for music
  source separation.
\newblock \emph{arXiv preprint:2109.05418}, 2021{\natexlab{a}}.

\bibitem[Kong et~al.(2021{\natexlab{b}})Kong, Ping, Huang, Zhao, and
  Catanzaro]{DiffWave}
Kong, Z., Ping, W., Huang, J., Zhao, K., and Catanzaro, B.
\newblock Diffwave: {A} versatile diffusion model for audio synthesis.
\newblock In \emph{{International Conference on Learning Representations}},
  2021{\natexlab{b}}.

\bibitem[Kreuk et~al.(2022)Kreuk, Synnaeve, Polyak, Singer, D{\'e}fossez,
  Copet, Parikh, Taigman, and Adi]{kreuk2022audiogen}
Kreuk, F., Synnaeve, G., Polyak, A., Singer, U., D{\'e}fossez, A., Copet, J.,
  Parikh, D., Taigman, Y., and Adi, Y.
\newblock Audiogen: Textually guided audio generation.
\newblock \emph{arXiv preprint:2209.15352}, 2022.

\bibitem[Kuleshov et~al.(2017)Kuleshov, Enam, and Ermon]{kuleshov2017audio}
Kuleshov, V., Enam, S.~Z., and Ermon, S.
\newblock Audio super resolution using neural networks.
\newblock \emph{arXiv:1708.00853}, 2017.

\bibitem[Lam et~al.(2022)Lam, Wang, Huang, Su, and Yu]{BDDM}
Lam, M., Wang, J., Huang, R., Su, D., and Yu, D.
\newblock Bilateral denoising diffusion models.
\newblock In \emph{{International Conference on Learning Representations}},
  2022.

\bibitem[Lee et~al.(2022)Lee, Kim, Shin, Tan, Liu, Meng, Qin, Chen, Yoon, and
  Liu]{PriorGrad}
Lee, S., Kim, H., Shin, C., Tan, X., Liu, C., Meng, Q., Qin, T., Chen, W.,
  Yoon, S., and Liu, T.
\newblock Priorgrad: Improving conditional denoising diffusion models with
  data-driven adaptive prior.
\newblock In \emph{{International Conference on Learning Representations}},
  2022.

\bibitem[Leng et~al.(2022)Leng, Chen, Guo, Liu, Chen, Tan, Mandic, He, Li, Qin,
  et~al.]{leng2022binauralgrad}
Leng, Y., Chen, Z., Guo, J., Liu, H., Chen, J., Tan, X., Mandic, D., He, L.,
  Li, X.-Y., Qin, T., et~al.
\newblock Binauralgrad: A two-stage conditional diffusion probabilistic model
  for binaural audio synthesis.
\newblock \emph{arXiv preprint:2205.14807}, 2022.

\bibitem[Liu et~al.(2021{\natexlab{a}})Liu, Kong, Tian, Zhao, Wang, Huang, and
  Wang]{liu2021voicefixer}
Liu, H., Kong, Q., Tian, Q., Zhao, Y., Wang, D., Huang, C., and Wang, Y.
\newblock Voicefixer: Toward general speech restoration with neural vocoder.
\newblock \emph{arXiv preprint:2109.13731}, 2021{\natexlab{a}}.

\bibitem[Liu et~al.(2022{\natexlab{a}})Liu, Choi, Liu, Kong, Tian, and
  Wang]{liu2022neural}
Liu, H., Choi, W., Liu, X., Kong, Q., Tian, Q., and Wang, D.
\newblock Neural vocoder is all you need for speech super-resolution.
\newblock \emph{arXiv preprint:2203.14941}, 2022{\natexlab{a}}.

\bibitem[Liu et~al.(2022{\natexlab{b}})Liu, Kong, Liu, Mei, Wang, and
  Plumbley]{liu2022ontology}
Liu, H., Kong, Q., Liu, X., Mei, X., Wang, W., and Plumbley, M.~D.
\newblock Ontology-aware learning and evaluation for audio tagging.
\newblock \emph{arXiv preprint:2211.12195}, 2022{\natexlab{b}}.

\bibitem[Liu et~al.(2022{\natexlab{c}})Liu, Liu, Kong, Wang, and
  Plumbley]{liu2022learning}
Liu, H., Liu, X., Kong, Q., Wang, W., and Plumbley, M.~D.
\newblock Learning the spectrogram temporal resolution for audio
  classification.
\newblock \emph{arXiv preprint:2210.01719}, 2022{\natexlab{c}}.

\bibitem[Liu et~al.(2022{\natexlab{d}})Liu, Li, Du, Torralba, and
  Tenenbaum]{CompositionalDDPM}
Liu, N., Li, S., Du, Y., Torralba, A., and Tenenbaum, J.~B.
\newblock Compositional visual generation with composable diffusion models.
\newblock In \emph{{European Conference on Computer Vision}},
  2022{\natexlab{d}}.

\bibitem[Liu et~al.(2021{\natexlab{b}})Liu, Iqbal, Zhao, Huang, Plumbley, and
  Wang]{liu2021conditional}
Liu, X., Iqbal, T., Zhao, J., Huang, Q., Plumbley, M.~D., and Wang, W.
\newblock Conditional sound generation using neural discrete time-frequency
  representation learning.
\newblock In \emph{IEEE International Workshop on Machine Learning for Signal
  Processing}, pp.\  1--6. IEEE, 2021{\natexlab{b}}.

\bibitem[Liu et~al.(2022{\natexlab{e}})Liu, Liu, Kong, Mei, Plumbley, and
  Wang]{liu2022simple}
Liu, X., Liu, H., Kong, Q., Mei, X., Plumbley, M.~D., and Wang, W.
\newblock Simple pooling front-ends for efficient audio classification.
\newblock \emph{arXiv preprint:2210.00943}, 2022{\natexlab{e}}.

\bibitem[Liu et~al.(2022{\natexlab{f}})Liu, Liu, Kong, Mei, Zhao, Huang,
  Plumbley, and Wang]{liu2022separate}
Liu, X., Liu, H., Kong, Q., Mei, X., Zhao, J., Huang, Q., Plumbley, M.~D., and
  Wang, W.
\newblock Separate what you describe: language-queried audio source separation.
\newblock \emph{arXiv preprint:2203.15147}, 2022{\natexlab{f}}.

\bibitem[Liu et~al.(2019)Liu, Ott, Goyal, Du, Joshi, Chen, Levy, Lewis,
  Zettlemoyer, and Stoyanov]{RoBERTa}
Liu, Y., Ott, M., Goyal, N., Du, J., Joshi, M., Chen, D., Levy, O., Lewis, M.,
  Zettlemoyer, L., and Stoyanov, V.
\newblock Roberta: {A} robustly optimized {BERT} pretraining approach.
\newblock \emph{arXiv preprint:1907.11692}, 2019.

\bibitem[Nichol \& Dhariwal(2021)Nichol and Dhariwal]{ImprovedDDPM}
Nichol, A. and Dhariwal, P.
\newblock Improved denoising diffusion probabilistic models.
\newblock In \emph{{International Conference on Machine Learning}}, 2021.

\bibitem[Nichol et~al.(2021)Nichol, Dhariwal, Ramesh, Shyam, Mishkin, McGrew,
  Sutskever, and Chen]{Glide}
Nichol, A., Dhariwal, P., Ramesh, A., Shyam, P., Mishkin, P., McGrew, B.,
  Sutskever, I., and Chen, M.
\newblock Glide: Towards photorealistic image generation and editing with
  text-guided diffusion models.
\newblock \emph{arXiv preprint:2112.10741}, 2021.

\bibitem[Oord et~al.(2016)Oord, Dieleman, Zen, Simonyan, Vinyals, Graves,
  Kalchbrenner, Senior, and Kavukcuoglu]{oord2016wavenet}
Oord, A. v.~d., Dieleman, S., Zen, H., Simonyan, K., Vinyals, O., Graves, A.,
  Kalchbrenner, N., Senior, A., and Kavukcuoglu, K.
\newblock {WaveNet}: A generative model for raw audio.
\newblock \emph{arXiv preprint:1609.03499}, 2016.

\bibitem[Pascual et~al.(2022)Pascual, Bhattacharya, Yeh, Pons, and
  Serr{\`a}]{pascual2022full}
Pascual, S., Bhattacharya, G., Yeh, C., Pons, J., and Serr{\`a}, J.
\newblock Full-band general audio synthesis with score-based diffusion.
\newblock \emph{arXiv preprint:2210.14661}, 2022.

\bibitem[Perez et~al.(2018)Perez, Strub, De~Vries, Dumoulin, and
  Courville]{perez2018film}
Perez, E., Strub, F., De~Vries, H., Dumoulin, V., and Courville, A.
\newblock Film: Visual reasoning with a general conditioning layer.
\newblock In \emph{Proceedings of the AAAI Conference on Artificial
  Intelligence}, volume~32, 2018.

\bibitem[Popov et~al.(2021)Popov, Vovk, Gogoryan, Sadekova, and
  Kudinov]{Grad-TTS}
Popov, V., Vovk, I., Gogoryan, V., Sadekova, T., and Kudinov, M.
\newblock Grad-tts: {A} diffusion probabilistic model for text-to-speech.
\newblock In \emph{{International Conference on Machine Learning}}, 2021.

\bibitem[Radford et~al.(2021)Radford, Kim, Hallacy, Ramesh, Goh, Agarwal,
  Sastry, Askell, Mishkin, Clark, Krueger, and Sutskever]{CLIP}
Radford, A., Kim, J.~W., Hallacy, C., Ramesh, A., Goh, G., Agarwal, S., Sastry,
  G., Askell, A., Mishkin, P., Clark, J., Krueger, G., and Sutskever, I.
\newblock Learning transferable visual models from natural language
  supervision.
\newblock In \emph{{International Conference on Machine Learning}}, 2021.

\bibitem[Raffel et~al.(2020)Raffel, Shazeer, Roberts, Lee, Narang, Matena,
  Zhou, Li, and Liu]{T5}
Raffel, C., Shazeer, N., Roberts, A., Lee, K., Narang, S., Matena, M., Zhou,
  Y., Li, W., and Liu, P.~J.
\newblock Exploring the limits of transfer learning with a unified text-to-text
  transformer.
\newblock \emph{Journal of Machine Learning Research}, 2020.

\bibitem[Ramesh et~al.(2022)Ramesh, Dhariwal, Nichol, Chu, and Chen]{DALLE2}
Ramesh, A., Dhariwal, P., Nichol, A., Chu, C., and Chen, M.
\newblock Hierarchical text-conditional image generation with clip latents.
\newblock \emph{arXiv preprint:2204.06125}, 2022.

\bibitem[Rombach et~al.(2022)Rombach, Blattmann, Lorenz, Esser, and
  Ommer]{rombach2022high}
Rombach, R., Blattmann, A., Lorenz, D., Esser, P., and Ommer, B.
\newblock High-resolution image synthesis with latent diffusion models.
\newblock In \emph{Proceedings of the IEEE/CVF Conference on Computer Vision
  and Pattern Recognition}, pp.\  10684--10695, 2022.

\bibitem[Saharia et~al.(2021)Saharia, Ho, Chan, Salimans, Fleet, and
  Norouzi]{ISRIR}
Saharia, C., Ho, J., Chan, W., Salimans, T., Fleet, D.~J., and Norouzi, M.
\newblock Image super-resolution via iterative refinement.
\newblock \emph{arXiv preprint:2104.07636}, 2021.

\bibitem[Saharia et~al.(2022)Saharia, Chan, Saxena, Li, Whang, Denton,
  Ghasemipour, Ayan, Mahdavi, Lopes, Salimans, Ho, Fleet, and Norouzi]{Imagen}
Saharia, C., Chan, W., Saxena, S., Li, L., Whang, J., Denton, E., Ghasemipour,
  S. K.~S., Ayan, B.~K., Mahdavi, S.~S., Lopes, R.~G., Salimans, T., Ho, J.,
  Fleet, D.~J., and Norouzi, M.
\newblock Photorealistic text-to-image diffusion models with deep language
  understanding.
\newblock \emph{arXiv preprint:2205.11487}, 2022.

\bibitem[Salamon et~al.(2014)Salamon, Jacoby, and Bello]{salamon2014dataset}
Salamon, J., Jacoby, C., and Bello, J.~P.
\newblock A dataset and taxonomy for urban sound research.
\newblock In \emph{Proceedings of the 22nd ACM international conference on
  Multimedia}, pp.\  1041--1044, 2014.

\bibitem[Schuhmann et~al.(2021)Schuhmann, Vencu, Beaumont, Kaczmarczyk, Mullis,
  Katta, Coombes, Jitsev, and Komatsuzaki]{schuhmann2021laion}
Schuhmann, C., Vencu, R., Beaumont, R., Kaczmarczyk, R., Mullis, C., Katta, A.,
  Coombes, T., Jitsev, J., and Komatsuzaki, A.
\newblock Laion-400m: Open dataset of clip-filtered 400 million image-text
  pairs.
\newblock \emph{arXiv preprint:2111.02114}, 2021.

\bibitem[Singer et~al.(2022)Singer, Polyak, Hayes, Yin, An, Zhang, Hu, Yang,
  Ashual, Gafni, Parikh, Gupta, and Taigman]{MakeAVideo}
Singer, U., Polyak, A., Hayes, T., Yin, X., An, J., Zhang, S., Hu, Q., Yang,
  H., Ashual, O., Gafni, O., Parikh, D., Gupta, S., and Taigman, Y.
\newblock Make-a-video: Text-to-video generation without text-video data.
\newblock \emph{arXiv preprint:2209.14792}, 2022.

\bibitem[Sinha et~al.(2021)Sinha, Song, Meng, and Ermon]{D2C}
Sinha, A., Song, J., Meng, C., and Ermon, S.
\newblock D2c: Diffusion-decoding models for few-shot conditional generation.
\newblock In \emph{{Conference on Neural Information Processing Systems}},
  2021.

\bibitem[Song et~al.(2020)Song, Meng, and Ermon]{song2020denoising}
Song, J., Meng, C., and Ermon, S.
\newblock Denoising diffusion implicit models.
\newblock \emph{arXiv preprint:2010.02502}, 2020.

\bibitem[Song et~al.(2021)Song, Sohl{-}Dickstein, Kingma, Kumar, Ermon, and
  Poole]{SGM}
Song, Y., Sohl{-}Dickstein, J., Kingma, D., Kumar, A., Ermon, S., and Poole, B.
\newblock Score-based generative modeling through stochastic differential
  equations.
\newblock In \emph{{International Conference on Learning Representations}},
  2021.

\bibitem[Tan et~al.(2022)Tan, Chen, Liu, Cong, Zhang, Liu, Wang, Leng, Yi, He,
  et~al.]{tan2022naturalspeech}
Tan, X., Chen, J., Liu, H., Cong, J., Zhang, C., Liu, Y., Wang, X., Leng, Y.,
  Yi, Y., He, L., et~al.
\newblock {NaturalSpeech}: End-to-end text to speech synthesis with human-level
  quality.
\newblock \emph{arXiv preprint:2205.04421}, 2022.

\bibitem[Vahdat et~al.(2021)Vahdat, Kreis, and Kautz]{LSGM}
Vahdat, A., Kreis, K., and Kautz, J.
\newblock Lsgm: Score-based generative modeling in latent space.
\newblock In \emph{{Conference on Neural Information Processing Systems}},
  2021.

\bibitem[Wang \& Wang(2021)Wang and Wang]{heming-towards-sr-wang2021towards}
Wang, H. and Wang, D.
\newblock Towards robust speech super-resolution.
\newblock \emph{IEEE/ACM Transactions on Audio, Speech, and Language
  Processing}, 29:\penalty0 2058--2066, 2021.

\bibitem[Wu et~al.(2022)Wu, Chen, Zhang, Hui, Berg-Kirkpatrick, and
  Dubnov]{wu2022large}
Wu, Y., Chen, K., Zhang, T., Hui, Y., Berg-Kirkpatrick, T., and Dubnov, S.
\newblock Large-scale contrastive language-audio pretraining with feature
  fusion and keyword-to-caption augmentation.
\newblock \emph{arXiv preprint:2211:06687}, 2022.

\bibitem[Yamagishi et~al.(2019)Yamagishi, Veaux, MacDonald,
  et~al.]{vctk-yamagishi2019cstr}
Yamagishi, J., Veaux, C., MacDonald, K., et~al.
\newblock {CSTR VCTK corpus}: English multi-speaker corpus for cstr voice
  cloning toolkit.
\newblock 2019.

\bibitem[Yang et~al.(2022)Yang, Yu, Wang, Wang, Weng, Zou, and
  Yu]{yang2022diffsound}
Yang, D., Yu, J., Wang, H., Wang, W., Weng, C., Zou, Y., and Yu, D.
\newblock Diffsound: Discrete diffusion model for text-to-sound generation.
\newblock \emph{arXiv preprint:2207.09983}, 2022.

\bibitem[{\.Z}elaszczyk \& Ma{\'n}dziuk(2022){\.Z}elaszczyk and
  Ma{\'n}dziuk]{zelaszczyk2022audio}
{\.Z}elaszczyk, M. and Ma{\'n}dziuk, J.
\newblock Audio-to-image cross-modal generation.
\newblock In \emph{International Joint Conference on Neural Networks}, pp.\
  1--8. IEEE, 2022.

\end{thebibliography}
\bibliographystyle{icml2021}

\newpage
\onecolumn
\section*{Appendix}
\label{sec:appendix}


\renewcommand{\thesubsection}{\Alph{subsection}}

\subsection{Contrastive Language-Audio Pretraining}
\label{app:CLAP}


We follow the pipeline of the contrastive language-audio pretraining (CLAP) models proposed by \cite{wu2022large} to capture the similarity between text and audio, and project them into joint latent space. The training dataset includes the currently largest public dataset LAION-Audio-$630$K, the AudioSet dataset whose text caption is augmented with keyword-to-caption\footnote{\url{https://github.com/gagan3012/keytotext}} by T$5$ model~\cite{T5}, the AudioCaps dataset and the Clotho dataset~\cite{Clotho}. The LAION-Audio-$630$K dataset contains $633,526$ language-audio pairs and $4325.39$ hours of audio samples. The AudioSet dataset contains $1,912,024$ pairs and $463.48$ hours of audio samples. The AudioCaps dataset contains $49,274$ pairs and $136.87$ hours of audio samples. The Clotho dataset contains $3,839$ pairs and $23.99$ hours of audio samples. These datasets contain various natural sounds, audio effects, music and human activity.

Given the audio sample $x$ and the text data $y$, we use an audio encoder and a text encoder to extract their embedding $\boldsymbol{E}^{x}\in \mathbb{R}^{L}$ and $\boldsymbol{E}^{y}\in \mathbb{R}^{L}$ respectively, where $L$ is set as $512$. We build the audio encoder based on HTSAT~\cite{HTSAT} and the text encoder based on RoBERTa~\cite{RoBERTa}. The symmetric cross-entropy loss used to train these contrastive encoders is:
\begin{align}
L_{s}&=\frac{1}{2D}\sum^{D}_{i=1}(l_{1} + l_{2}),\\
l_{1}&=\log\frac{\exp(\boldsymbol{E}^{x}_{i}\cdot \boldsymbol{E}^{y}_{i}/\tau)}{\sum^{N}_{i=1}\exp(\boldsymbol{E}^{x}_{i}\cdot \boldsymbol{E}^{y}_{j}/\tau)},\\
l_{2}&=\log\frac{\exp(\boldsymbol{E}^{y}_{i}\cdot \boldsymbol{E}^{x}_{i}/\tau)}{\sum^{N}_{i=1}\exp(\boldsymbol{E}^{y}_{i}\cdot \boldsymbol{E}^{x}_{j}/\tau)}),
\end{align}
where $\tau$ is a learnable temperature parameter and $D$ is the batch size. 

\subsection{Latent Diffusion Model}
\label{app:LDMArchitecture}

We adopt the UNet backbone of StableDiffusion~\cite{rombach2022high} as the basic architecture of LDM for AudioLDM. As shown in Equation~\ref{singlereversestep}, the UNet model is conditioned on both the time step $t$ and the CLAP embedding $\boldsymbol{E}$. We map the time step into a one-dimensional embedding and then concatenate it with $\boldsymbol{E}$ as conditioning information. Since our condition vector is only one-dimensional, we do not use the cross-attention mechanism in StableDiffusion for conditioning. Instead, we directly use the feature-wise linear modulation layer~\cite{perez2018film} to merge conditioning information with the feature map of the UNet convolution block. The UNet backbone we use has four encoder blocks, a middle block, and four decoder blocks. With a basic channel number of $c_{u}$, the channel dimensions of encoder blocks are $[c_{u}, 2c_{u}, 3c_{u}, 5c_{u}]$. The channel dimensions of decoder blocks are the reverse of encoder blocks, and the channel of the middle block has $5c_{u}$ dimensions. We add an attention block in the last three encoder blocks and the first three decoder blocks. Specifically, we add two multi-head self-attention layers with a fully-connected layer in the middle as the attention block. The number of heads is determined by dividing the embedding dimension of the attention block with a parameter $c_h$. We set AudioLDM-S and AudioLDM-L with $c_{u}\vequalsignnospace128, c_h\vequalsignnospace32$, and $c_{u}\vequalsignnospace256, c_h\vequalsignnospace64$, respectively. In the forward process, we use $N=1000$ steps. A linear noise schedule from $\beta_{1}=0.0015$ to $\beta_{N}=0.0195$ is used. In sampling, we employ the DDIM~\cite{song2020denoising} sampler with $200$ sampling steps. For classifier-free guidance, a guidance scale $w$ of $2.0$ is used in Equation \ref{mixup}.

\subsection{Variational Autoencoder}
\label{app:VAE}

\begin{figure}[htbp]
    \centering
    \includegraphics[width=0.8\linewidth]{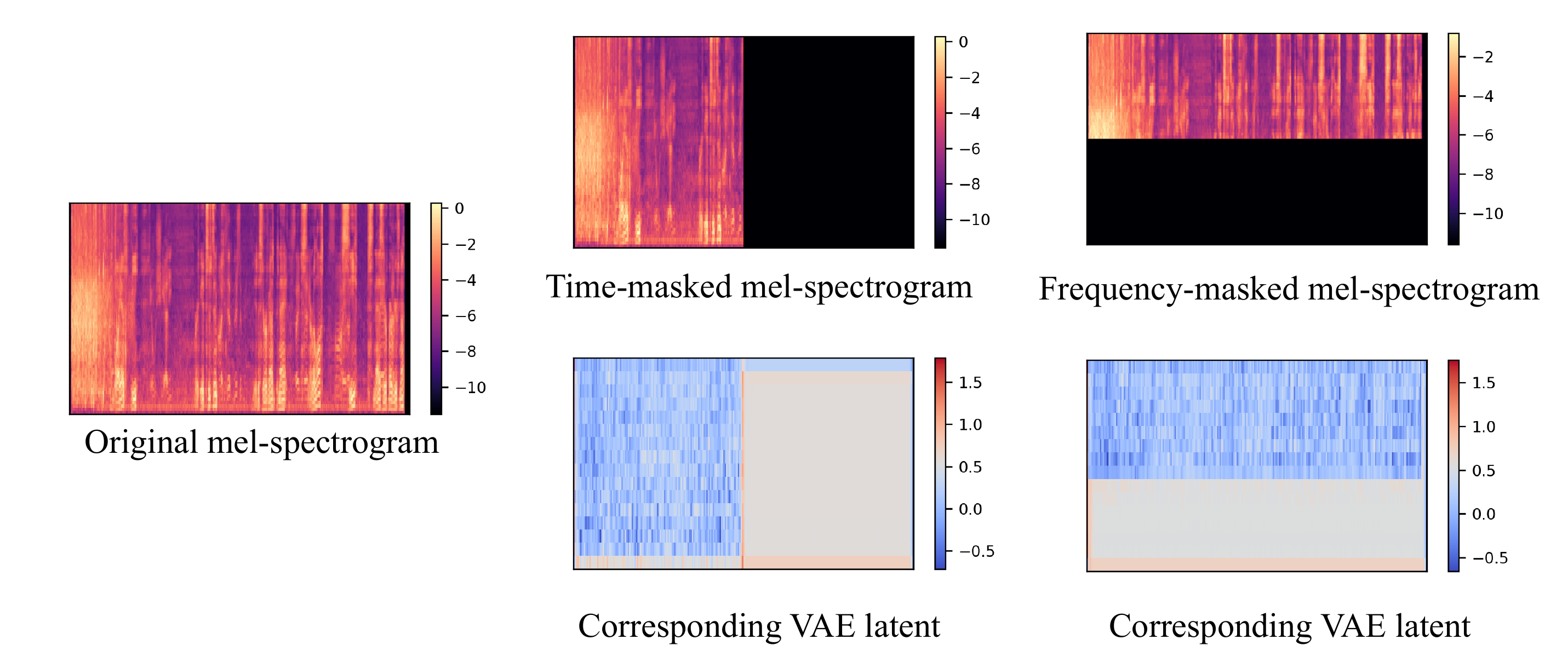}
    \caption{Visualization of the time-frequency-masked spectrogram and their corresponding VAE latent. This figure shows VAE encoder roughly preserves the spatial correspondancy between the spectrogram and the latent.}
    \label{fig:demo-spatial-correspondancy}
\end{figure}

We compress the mel-spectrogram $\boldsymbol{X}\in \mathbb{R}^{T\times F}$ of $x$ into a small continuous space $\boldsymbol{z}\in \mathbb{R}^{C\times \frac{T}{r}\times \frac{F}{r}}$ with a convolutional VAE, where $T$ and $F$ is the time and frequency dimension size respectively, $C$ is the channel number of the latent encoding, and $r$ is the compression level (downsampling ratio) of latent space. Both the encoder $\mathcal{E}(\cdot)$ and the decoder $\mathcal{D}(\cdot)$ are composed of stacked convolutional modules. In this way, VAE encoder could preserve the spatial correspondancy between mel-spectrogram and latent space, as it is shown in Figure~\ref{fig:demo-spatial-correspondancy}. Each module is formed by ResNet blocks~\cite{kong2021decoupling} which are made up of convolutional layers and residual connections. The encoding $\boldsymbol{z}$ will be evenly split into two parts, $\boldsymbol{z}_{\boldsymbol{\mu}}$ and $\boldsymbol{z}_{\boldsymbol{\sigma}}$, with shape $(\frac{C}{2}, \frac{T}{r}, \frac{F}{r})$, representing the mean and variance of the VAE latent space. The input of the decoder is a stochastic encoding $\boldsymbol{\hat{z}}=\boldsymbol{\hat{z}}_{\boldsymbol{\mu}}+\boldsymbol{\hat{z}}_{\boldsymbol{\sigma}} \cdot \boldsymbol{\epsilon}, \boldsymbol{\epsilon} \sim \mathcal{N}(0, I)$. During generation, the decoder will be used to reconstruct the mel-spectrogram given the generated latent representations.

We employ three loss functions in our training objective: the mel-spectrogram reconstruction loss, adversarial losses, and a gaussian constraint loss. The reconstruction loss calculates the mean absolute error between the input sample $\boldsymbol{X}\in \mathbb{R}^{T\times F}$ and the reconstructed mel-spectrogram $\hat{\boldsymbol{X}}\in \mathbb{R}^{T\times F}$. The adversarial losses are employed to enhance the reconstruction quality. Specifically, we adopt the PatchGAN~\citep{isola2017image} as our discriminator, which will divide the input image into small patches and predict whether each patch is real or fake by outputting a matrix of logits. 
The PatchGAN discriminator is trained to maximize the logits of correctly identifying real patches while minimizing the logits of incorrectly identifying fake patches. We also apply the gaussian constraint on the latent space of VAE. By enforcing a gaussian constraint on the latent space, the VAE is encouraged to learn a continuous, structured latent space, rather than a disorganized one. This can help the VAE to better capture the underlying structure of the data, which can result in more stabilized and accurate reconstructions~\cite{kingma2013auto}.

We train our VAE using the Adam optimizer~\cite{kingma2014adam} with a learning rate of $4.5\times 10^{-6}$ and a batch size of six. The audio data we use includes AudioSet, AudioCaps, Freesound, and BBC SFX. We perform experiments with three compression-level settings $r\vequalsignnospace4,8,16$, for which the latent channels are $C=8,16,32$, respectively.  VAEs in all three settings are trained with at least $1.5$M steps on a single NVIDIA RTX 3090 GPU. To stabilize training, we do not apply the adversarial loss in the first $50$K training steps. We apply the mixup~\cite{kong2020panns} strategy for data augmentation.

Table~\ref{tab: vae-reconstruct} shows the reconstruction performance of our VAE model with different values of $r$. All three settings achieve comparable metrics score with the \textit{GT Mel + Vocoder} setting, indicating the autoencoder can perform reliable mel-spectrogram encoding and decoding.


\begin{table}[htbp]
\small
\centering
\begin{tabular}{cccccc}
\toprule
        Setting        & PSNR$\uparrow$  & SSIM$\uparrow$ & FD$\downarrow$   & IS$\uparrow$   & KL$\downarrow$   \\
\midrule
GT Mel + Vocoder & $25.41$ & $0.86$ & $8.76$ & $10.71$ & $0.23$ \\
\midrule
Compression$_{r=4}$     & $25.38$ & $0.86$ & $9.02$ & $10.67$ & $0.23$ \\
Compression$_{r=8}$     & $25.14$ & $0.84$ & $9.68$ & $10.50$ & $0.25$ \\
Compression$_{r=16}$    & $24.87$ & $0.82$ & $9.90$ & $9.84$ & $0.29$ \\
\bottomrule
\end{tabular}
\caption{The objective metrics of VAE reconstruction performance with different compression level $r$ on the AudioSet evaluation set.}
\label{tab: vae-reconstruct}
\end{table}

\subsection{Vocoder}
\label{app:HiFi-GAN}

In this work, we employ HiFi-GAN~\cite{kong2020hifi} as a vocoder, which is widely used for speech waveform generation. It contains two sets of discriminators, a multi-period discriminator, and a multi-scale discriminator, to enhance the perceptual quality. To synthesize the audio waveform, we train it on the AudioSet dataset. For the input samples at the sampling rate of $16,000$Hz, we extract $64$ bands mel-spectrogram. Then we follow the default settings of HiFi-GAN V$1$. The window, FFT, and hop size are set to $1024$, $1024$, and $160$. The $f_{\text{min}}$ and $f_{\text{max}}$ are set as $0$ and $8000$. We use the AdamW optimizer with $0.8$ and $0.99$. The learning rate starts from $2\times 10^{-4}$ and a learning rate decay of $0.999$ is used. We use a batch size of $96$ and train the model with $6$ NVIDIA $3090$ GPUs. We release this pretrained vocoder in our open-source implementation.

\subsection{Experiment Details}
\label{app:TrainingDetails}

\textbf{Data Processing} The duration of the audio samples in AudioSet and AudioCaps is $10$ seconds, while it is much longer in FreeSound and BBC SFX datasets. To avoid overusing the data from long audio, which usually have repeated sound, we only use the first thirty seconds of the audio in both the FreeSound and BBC SFX datasets and segment them into ten-second long audio files. Finally, we have in total $3,302,553$ ten-seconds audio samples for model training. It should be noted that even if some datasets, e.g., AudioCaps and BBC SFX, have text captions for the audio, we do not utilize them during the training of LDMs. We only use the audio samples for training. We resample all the datasets into $16$kHz sampling rate and mono format, and all samples are padded to $10$ seconds. 

\textbf{Configuration} For each LDM model, we use the compression level $r\vequalsignnospace4$ as the default setting. Then, we train AudioLDM-S and AudioLDM-L for $0.6$M steps on a \textbf{single GPU}, NVIDIA RTX $3090$, with the batch size of $5$ and $8$, respectively. The learning rate is set as $3\times 10^{-5}$. The AudioLDM-L-Full is trained for $1.5$M steps on one NVIDIA A$100$ with a batch size of $8$. The learning rate is $10^{-5}$. For better performance on AudioCaps, we further fine-tune AudioLDM-L-Full on AudioCaps for $0.25$M steps before evaluation. It should be noted that we limit our batch size because of the scarcity of GPU. However, this potentially restricts the performance of AudioLDM models. In comparison, DiffSound uses $32$ NVIDIA V$100$ GPUs for model training with a batch size of $16$ on each GPU. AudioGen utilizes $64$ A$100$ GPUs with a batch size of $256$. 

\textbf{Human evaluation} We construct the  dataset for human subjective evaluation with $70$ randomly selected samples where $30$ audios are from AudioCaps, $30$ audios are from AudioSet, and $10$ randomly selected real recordings, which we will refer to as spam cases. Therefore, each model should generate $60$ audio samples given the corresponding text descriptions. We gather the output from models in one folder and anonymize them with random identifiers. An example questionnaire is shown in Table~\ref{tab:questionniare}. The participant will need to fill in the last two columns for each audio file given the text description. Our final result shows that all the human raters have an average score above $90$ on the spam cases. Hence, their evaluation result is considered reliable.

\begin{table}[htbp]
\scriptsize
\centering
\begin{tabular}{cccc}
\toprule
File name &
  Text description &
  Overall impression (1-100) &
  Relation to the text description   (1-100) \\
\midrule
random\_name\_108029.wav &
  A man talking followed by lights scrapping on a wooden surface &
  80 &
  90 \\
\midrule
random\_name\_108436.wav & Bicycle Music Skateboard   Vehicle          & 70 & 80 \\
\midrule
random\_name\_116883.wav & A power tool drilling as rock   music plays & 90 & 95 \\
\midrule
... & ... & ... & ... \\
\bottomrule
\end{tabular}
\caption{Example questionnaire for human evaluation. The participant will need to fill in the last two columns.}
\label{tab:questionniare}
\end{table}

\subsection{The Effect of Finetuning}

\begin{table}[tbp]
\centering
\small
\begin{tabular}{cccc|cccc}
\toprule
    Model    & Text Data & Use CLAP  & Finetuned & FD~$\downarrow$  & IS~$\uparrow$   & KL~$\downarrow$ & FAD~$\downarrow$   \\
\midrule
AudioLDM-S-Full-Roberta   & \cmark   & \xmark      & \xmark  & $34.28$  & $3.53$ & $3.44$ & $6.96$ \\
AudioLDM-S-Full-Roberta   & \cmark   & \xmark      & \cmark  & $32.13$  & $4.02$ & $3.25$ & $5.89$ \\
AudioLDM-S-Full & \xmark & \cmark  & \xmark & $24.13$  & $6.68$ & $2.36$ & $4.94$ \\
AudioLDM-S-Full & \xmark & \cmark  & \cmark & $23.47$  & $7.57$ & $1.98$ & $2.32$ \\
AudioLDM-L-Full & \xmark & \cmark  & \xmark & $23.51$  & $7.11$ & $2.19$ & $4.19$  \\
AudioLDM-L-Full & \xmark & \cmark  & \cmark & $23.31$  & $8.13$ & $1.59$ & $1.96$  \\
\bottomrule
\end{tabular}
\caption{The comparison between fine-tuned and non-finetuned models on the AudioCaps evaluation set.}
\label{tab: finetune-non-finetune-comparison}
\end{table}

Table~\ref{tab: finetune-non-finetune-comparison} compares the results obtained with and without fine-tuning on the evaluation set. We observe an improvement in various evaluation metrics, which is expected since the training set of AudioCaps has a distribution that is similar to the evaluation set. However, it is important to note that higher performance on the limited distribution of the evaluation set may not necessarily indicate better performance overall. A model that can generate broader distributions of audio may perform worse on the evaluation set, even though it may have better generalization capabilities. Future work in audio generation can focus on building an evaluation protocal that is more aligned with human perceptions.

\subsection{Computation Efficiency Comparison}

\begin{figure}[htbp]
  \centering
  \subfigure[Different batch sizes]{\includegraphics[width=0.3\textwidth]{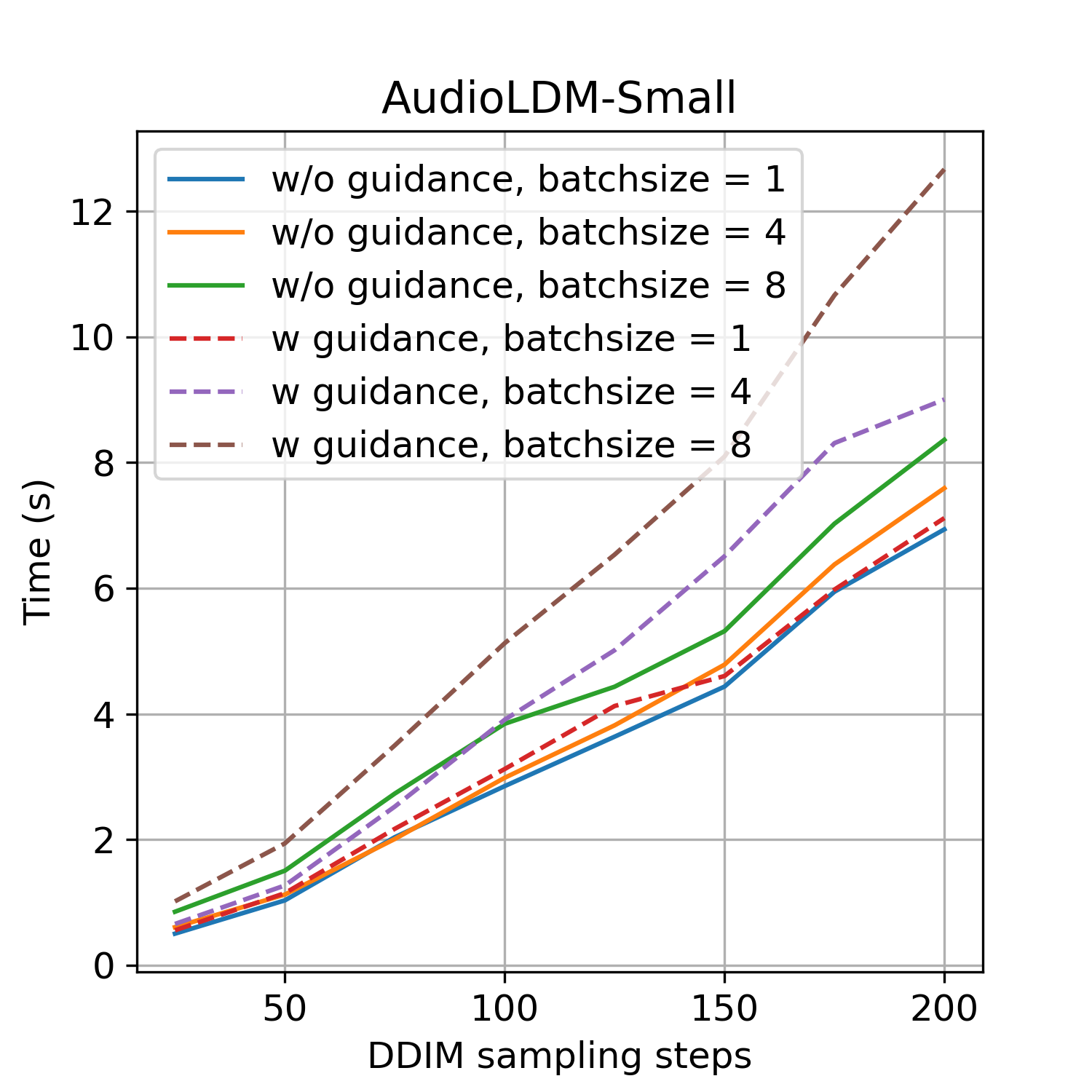}\label{fig:sub-audioldm-s}}
  \subfigure[Different sampling steps]{\includegraphics[width=0.3\textwidth]{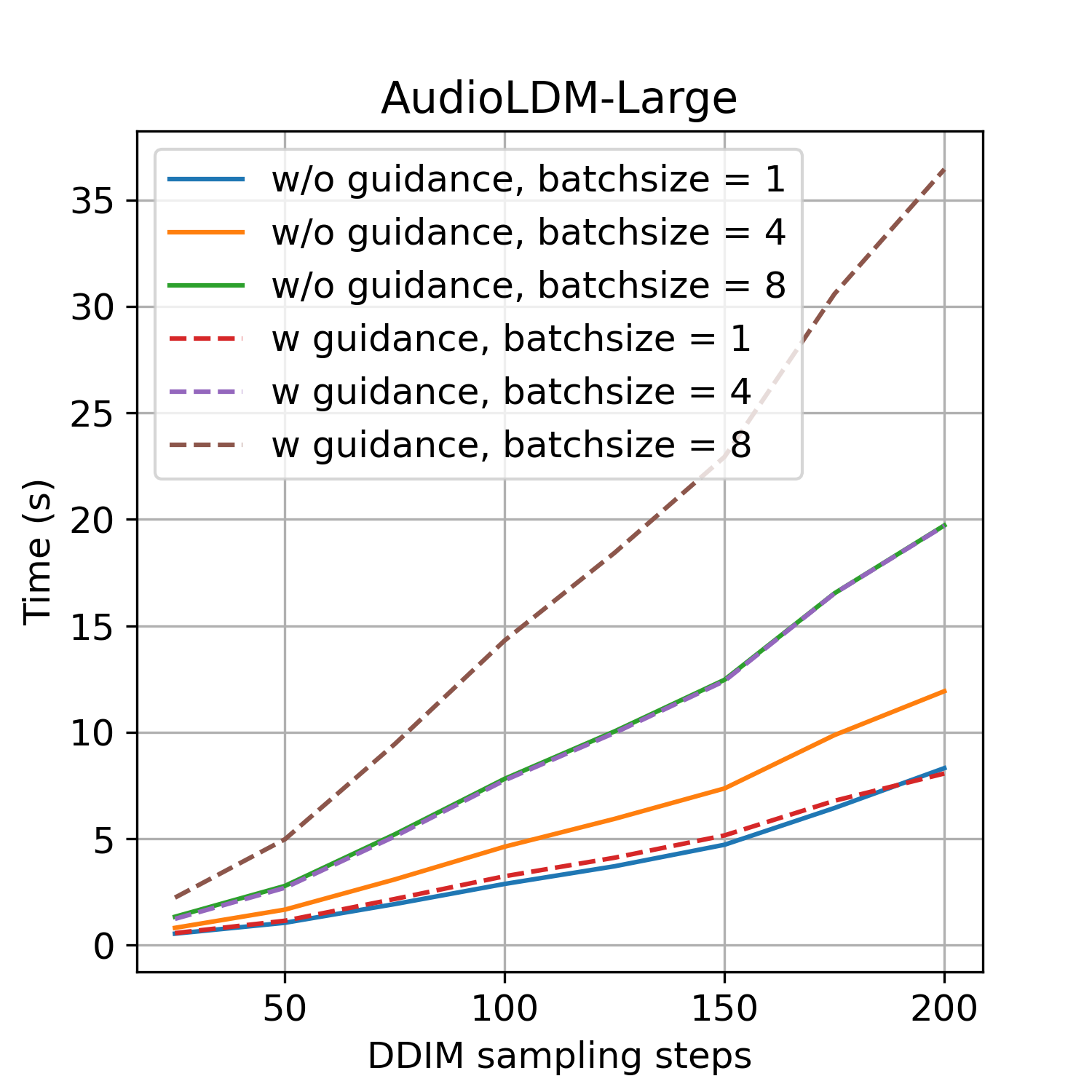}\label{fig:sub-audioldm-l}}
  \subfigure[With/Without classifier-free guidance]{\includegraphics[width=0.3\textwidth]{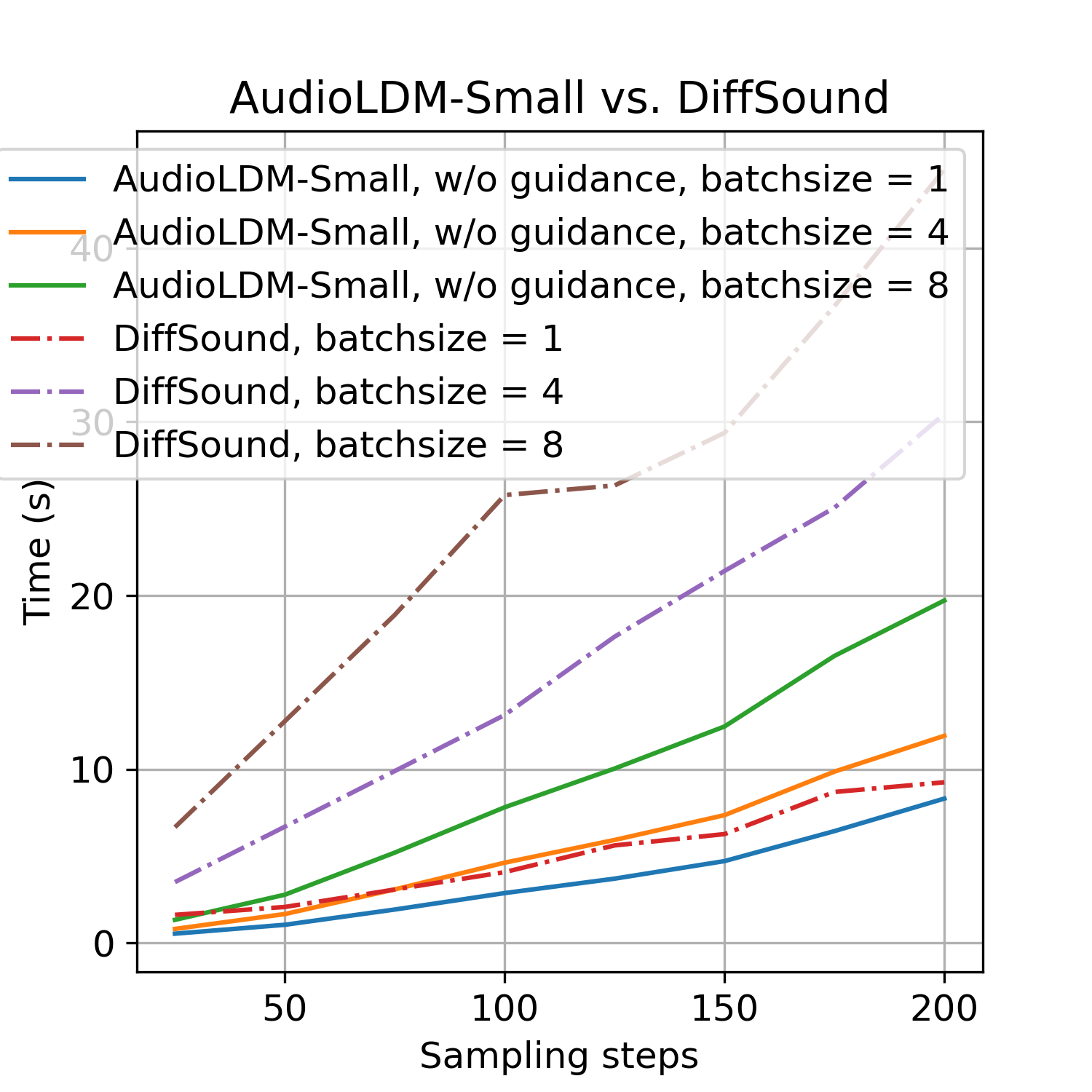}\label{sub-audioldm-timing}}
  \caption{These three figures show the time cost when generating ten seconds of audio, measured on a single A100 GPU.}
  \label{fig:speed-comparison}
\end{figure}

As shown in Figure~\ref{fig:sub-audioldm-s}, AudioLDM-S can generate eight ten-second-long audios within ten seconds without classifier-free guidance. With classifier-free guidance, AudioLDM-Small can generate eight ten-second-long audios with $150$ DDIM steps. Figure~\ref{sub-audioldm-timing} shows our model is faster than the DiffSound on different batch sizes. Our model can generate eight ten-second-long audios with $20$ seconds while DiffSound needs more than $40$ seconds. Since AudioGen has not been open-sourced yet, we did not perform a speed comparison with AudioGen.

\subsection{Limitations}

There are several limitations to our study that warrant further investigation in future work. For example, the sampling rate of our model is still insufficient, especially for the generation of music. Exploring higher-fidelity sampling rates such as 32 kHz or 48 kHz could improve the quality of the generated audio. Also, all the modules in AudioLDM are trained separately, which may result in misalignment between different modules. For instance, the latent space learned by VAE may not be optimal for the latent diffusion model. Future work can explore approaches to better align the different modules, such as end-to-end fine-tuning.

The possible negative impact of our method might be the abuse of our technology or released models, e.g., generating fake audio effects to provide misleading information. Moreover, sensitive text content should be restricted in future work to prevent the creation of harmful audio content.

\newpage
\subsection{Demos}
\label{app:demos}

\textbf{Audio Style Transfer}

\begin{figure}[htbp]
    \centering
    \includegraphics[width=0.75\linewidth]{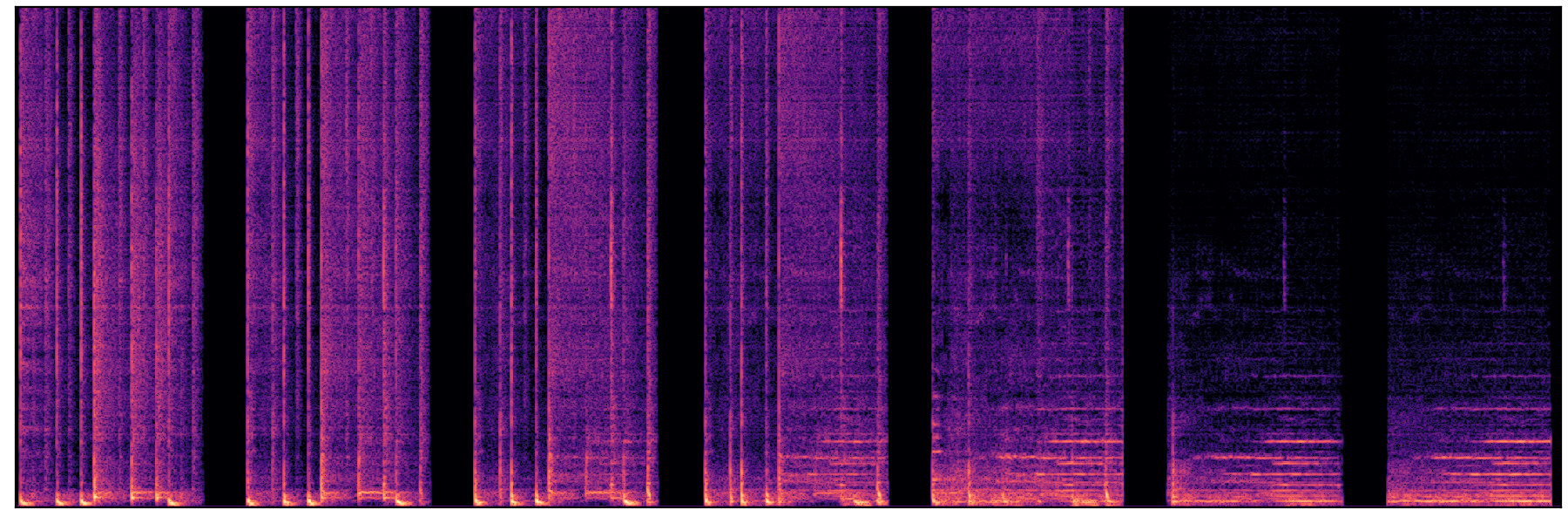}
    \caption{Audio style transfer from \textbf{drum beats} to \textbf{ambient music}.}
    \label{fig:demo-style-transfer-1}
\end{figure}

\begin{figure}[htbp]
    \centering
    \includegraphics[width=0.75\linewidth]{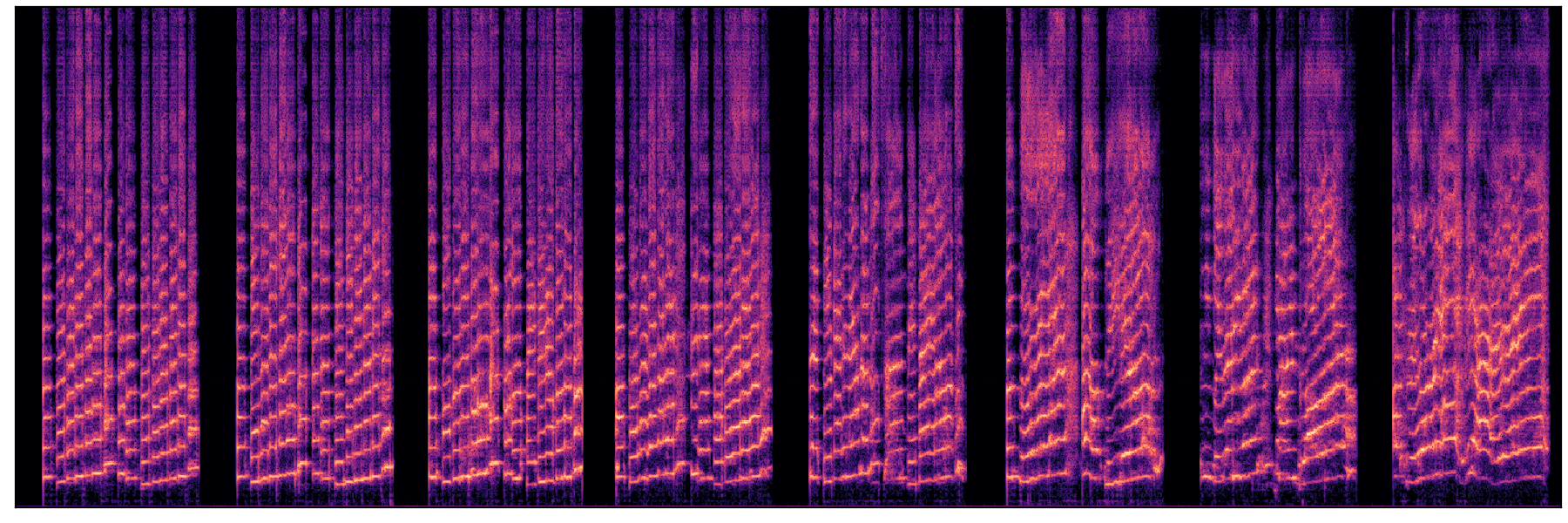}
    \caption{Audio style transfer from \textbf{trumpet} to \textbf{children singing}.}
    \label{fig:demo-style-transfer-2}
\end{figure}

\begin{figure}[htbp]
    \centering
    \includegraphics[width=0.75\linewidth]{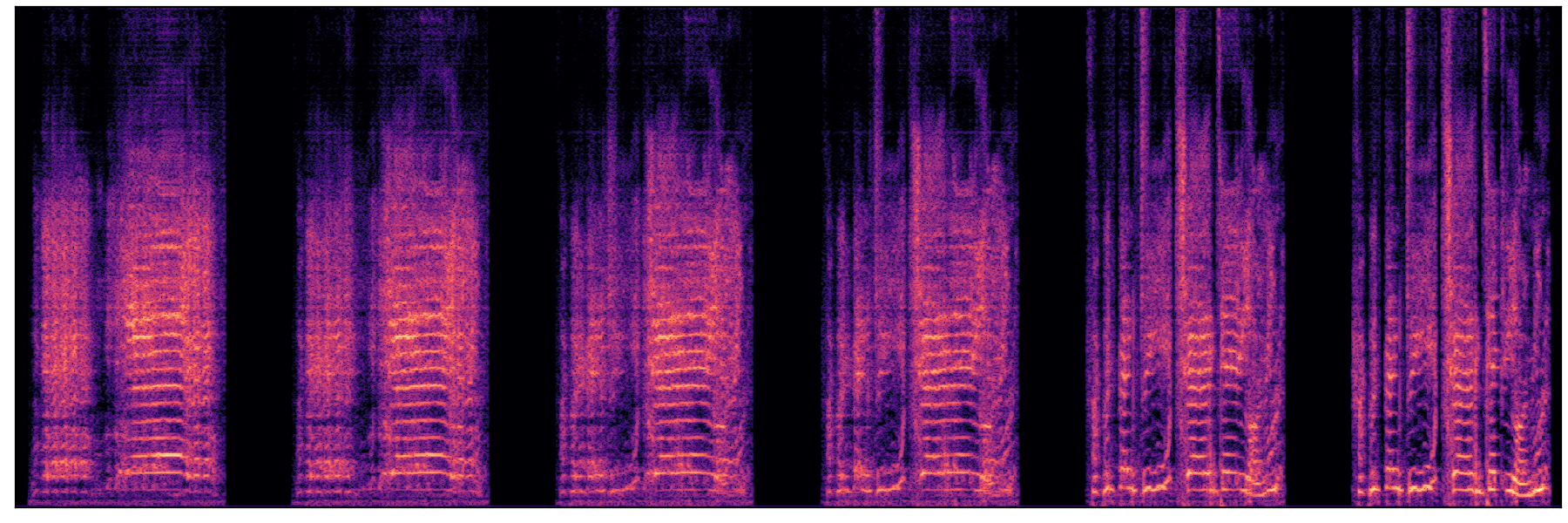}
    \caption{Audio style transfer from \textbf{sheep vocalization} to \textbf{narration, monologue}.}
    \label{fig:demo-style-transfer-3}
\end{figure}

We show three examples of zero-shot audio style transfer with AudioLDM-S, using the developed shallow reverse process~(see Equation~\ref{shallowreverse}). In Figure~\ref{fig:demo-style-transfer-1}, we show the transfer from \textbf{drum beats} to \textbf{ambient music}. From left to right, we show the source audio sample drum beats, and the six generated samples guided by text prompt \textbf{ambient music} with different starting points $n_{0}$. Given a smaller $n_{0}$~(i.e., the left part of the figure), the generated sample is similar to drum beats, while when we set $n_{0}=0.8\times N$ for the last sample, the generated sample will be aligned with the text input \textbf{ambient music}. Similarly, we show the source audio \textbf{trumpt}, and the seven generated samples guided by text prompt \textbf{children singing} in Figure~\ref{fig:demo-style-transfer-2}. We show the source audio \textbf{sheep vocalization}, and the five generated samples guided by text prompt \textbf{narration, monologue} in Figure~\ref{fig:demo-style-transfer-3}.

\vspace{0.5cm}

\newpage

\textbf{Audio Super-Resolution}

\begin{figure}[htbp]
    \centering
    \includegraphics[width=0.85\linewidth]{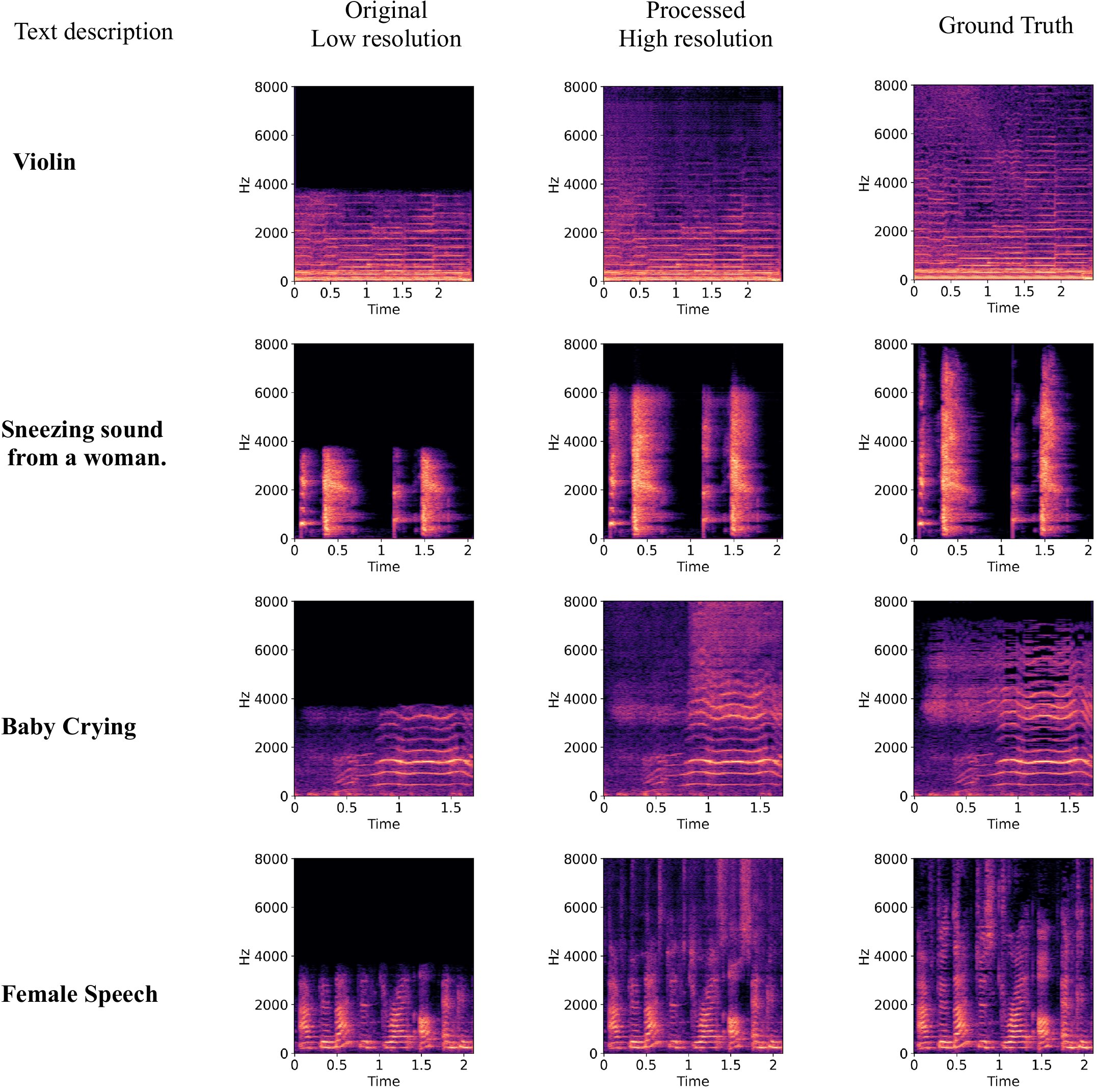}
    \caption{The examples of zero-shot audio super-resolution with AudioLDM-S.}
    \label{fig:demo-super-resolution}
\end{figure}

In Figure~\ref{fig:demo-super-resolution}, we show four cases of zero-shot audio super-resolution with AudioLDM-S: $1$) \textbf{violin}, $2$) \textbf{sneezing sound from a woman}, $3$) \textbf{baby crying}, and $4$) \textbf{female speech}. The sampling rate of input samples (left) is $8$~kHz, and that of generated samples (middle) and ground-truth samples (right) is $16$~kHz. Our visualization shows we can retain the ground-truth observation in the low-frequency part (below $8$~kHz), while generating the high-frequency missing part (from $8$~kHz to $16$~kHz) with pretrained AudioLDM-S. The generated high-frequency information is consistent with the low-frequency observation.  

\vspace{0.5cm}

\newpage

\textbf{Audio Inpainting}

\begin{figure}[htbp]
    \centering
    \includegraphics[width=1.0\linewidth]{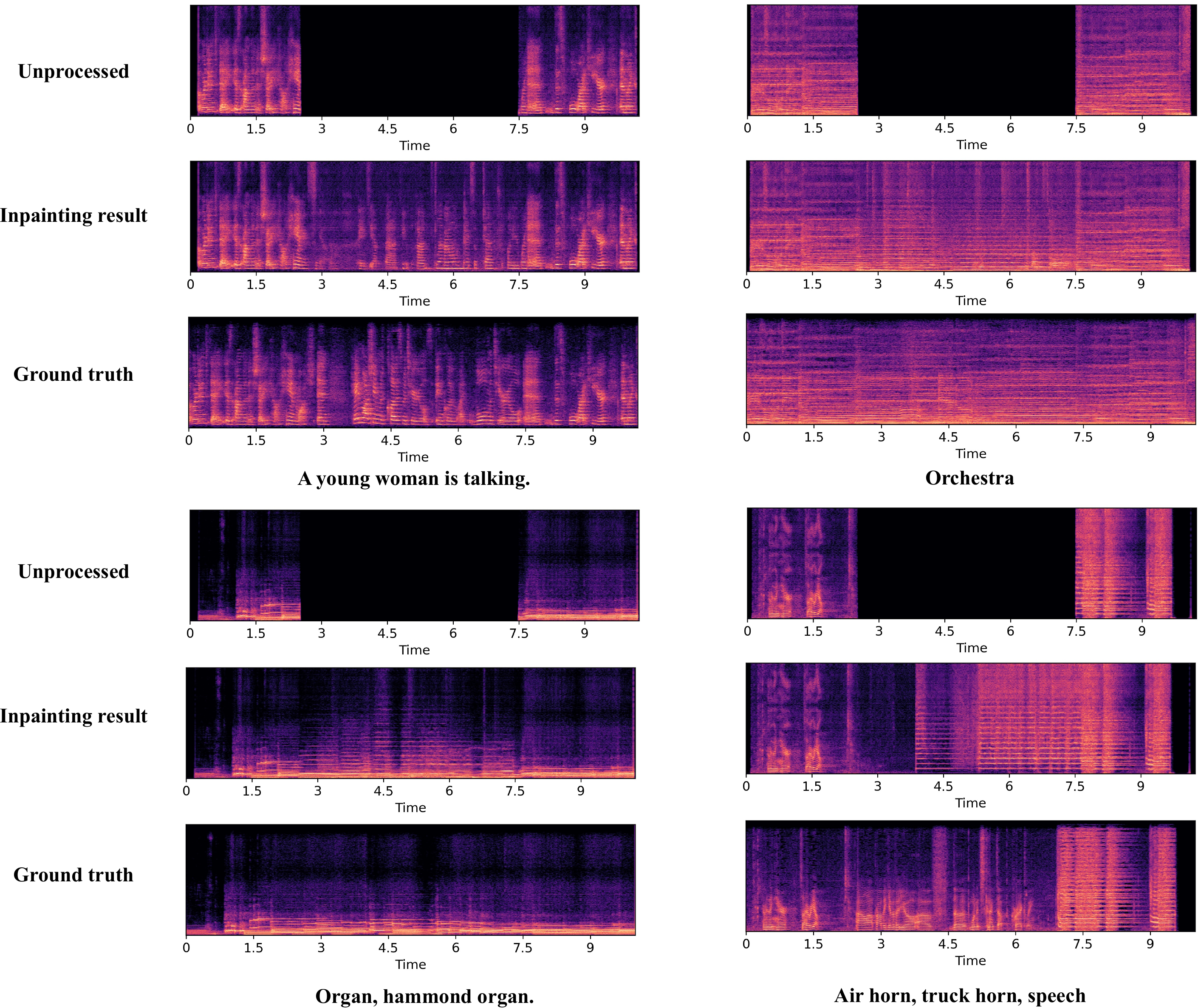}
    \caption{The examples of zero-shot audio inpainting with AudioLDM-S.}
    \label{fig:demo-inpainting}
\vspace{0.5cm}
\end{figure}

In Figure~\ref{fig:demo-inpainting}, we show four samples of zero-shot audio inpainting with AudioLDM-S. The time length of each audio sample is $10$ seconds. In the \textbf{unprocessed} part, we remove the content between $2.5$ and $7.5$ seconds from the ground-truth sample as the input of inpainting. In the \textbf{inpainting result} part, we show the generated samples guided by the same text prompt of the ground-truth sample. In the \textbf{ground truth} part, we show the ground-truth sample for comparison.

\vspace{0.5cm}

\begin{figure}[htbp]
    \centering
    \includegraphics[width=1.0\linewidth]{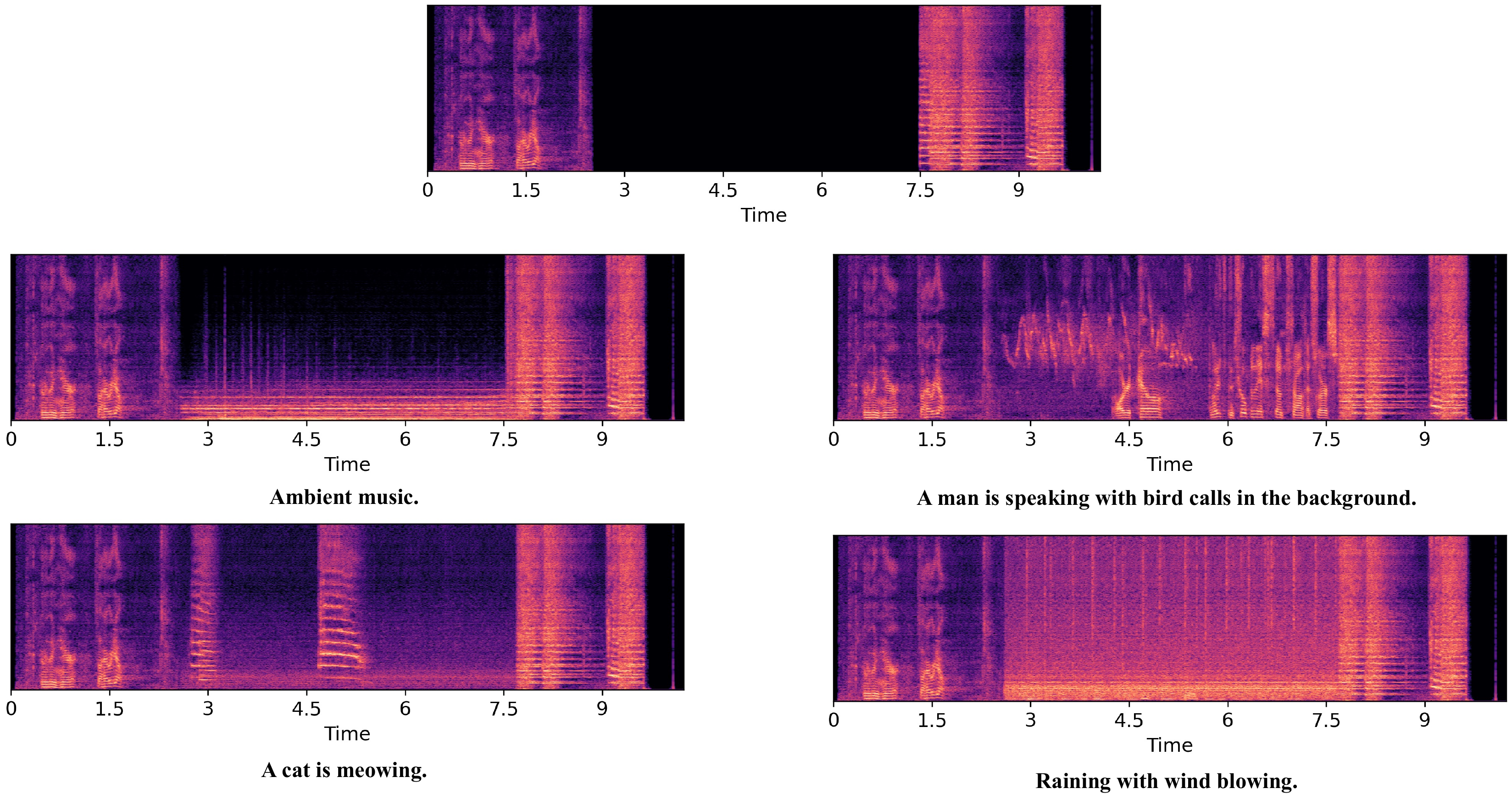}
    \caption{The example of zero-shot audio inpainting with AudioLDM-S given different text prompts.}
    \label{fig:demo-inpainting-prompt}
\vspace{0.5cm}
\end{figure}

In Figure~\ref{fig:demo-inpainting-prompt}, we use one sample to demonstrate the audio inpainting guided by different text prompts. Given the observed audio signal shown in the top row, we guide the inpainting process with four different text prompts: $1$) \textbf{ambient music}; $2$) \textbf{a man is speaking with bird calls in the background}; $3$) \textbf{a cat is meowing}; $4$) \textbf{raining with wind blowing}. As can be seen, the observed audio signal is preserved in each generated sample, while the generated content can be controlled by text input.   

\newpage

\vspace{0.5cm}

\textbf{Environment Control}

\begin{figure}[H]
    \centering
    \includegraphics[width=1.0\linewidth]{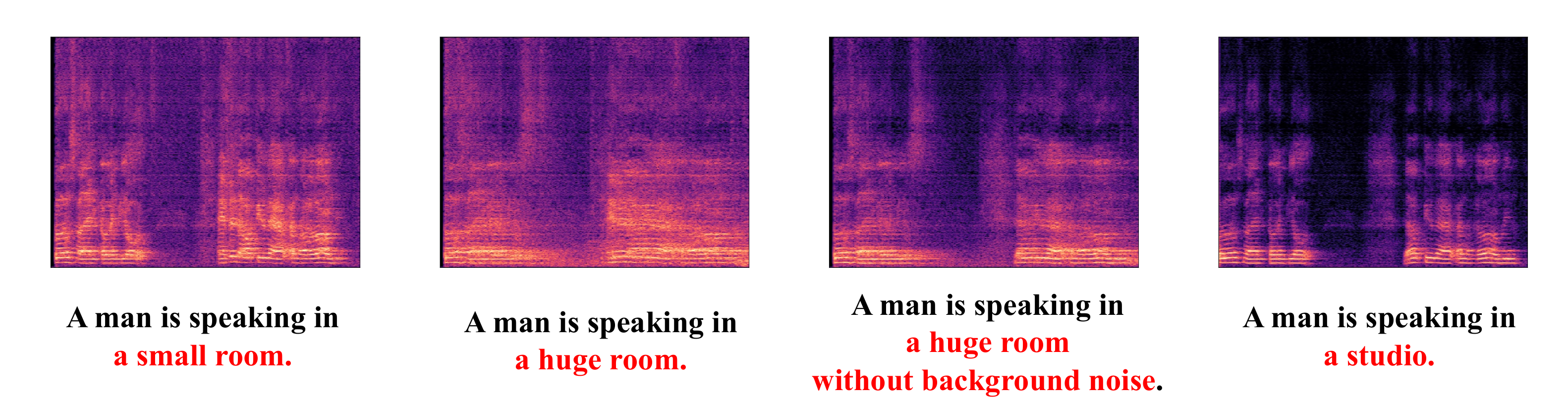}
    \caption{The examples of controlling acoustic environment with AudioLDM-S.}
    \label{fig:demo-control-speech}
\end{figure}

In Figure~\ref{fig:demo-control-speech}, we demonstrate that AudioLDM can control the acoustic environment of generated samples with a text description. The four samples are generated with the same random seed, but with different text prompts. Their common text information is \textbf{``A man is speaking in''}, while the specific text information describes the acoustic environment as \textbf{``a small room''}, \textbf{``a huge room''}, \textbf{``a huge room without background noise''}, and \textbf{``a studio''}. These samples show the ability of AudioLDM to capture the fine-grained text description about the acoustic environment, and control the corresponding effects on audio samples, such as reverberation or background noise.   
\vspace{0.5cm}

\newpage

\textbf{Music Control}

\begin{figure}[H]
    \centering
    \includegraphics[width=0.7\linewidth]{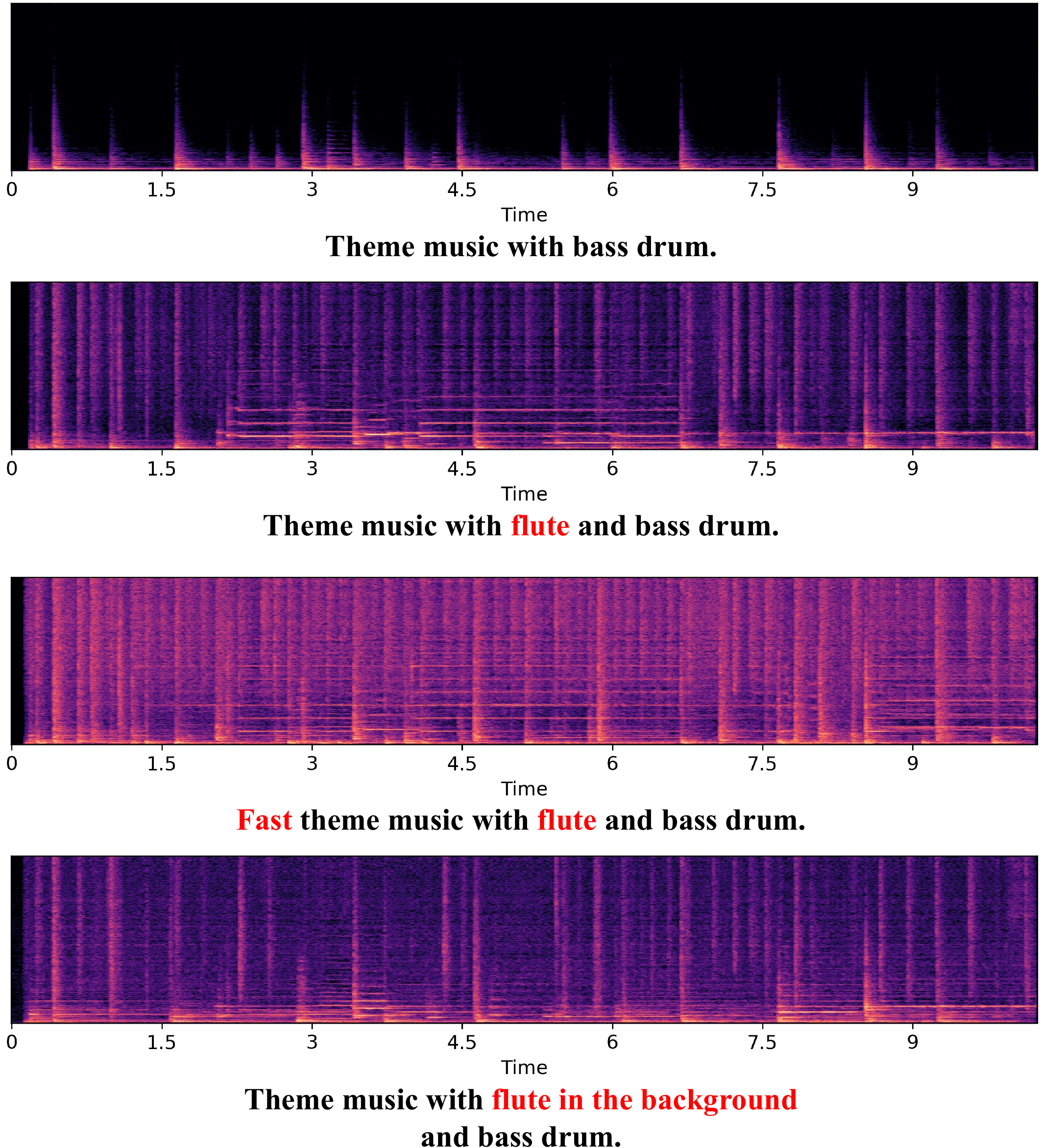}
    \caption{The examples of controlling music characteristics with AudioLDM-S.}
    \label{fig:demo-control-music}
\vspace{0.5cm}
\end{figure}

In Figure~\ref{fig:demo-control-music}, we show the generated music samples when we control the music characteristics with text input. The first sample is generated by \textbf{``Theme music with bass drum''}. Then, we add specific text information \textbf{``flute''}, \textbf{``fast, flute''}, or \textbf{``flute in the background''}, to change the text input. The corresponding variations can be seen in generated mel-spectrograms. We use these samples to demonstrate the ability of AudioLDM to add new musical instruments to music samples, tune the speed of music, and control the foreground-background relations. 

\vspace{0.8cm}

\newpage    

\textbf{Pitch Control}

\begin{figure}[H]
    \centering
    \includegraphics[width=0.5\linewidth]{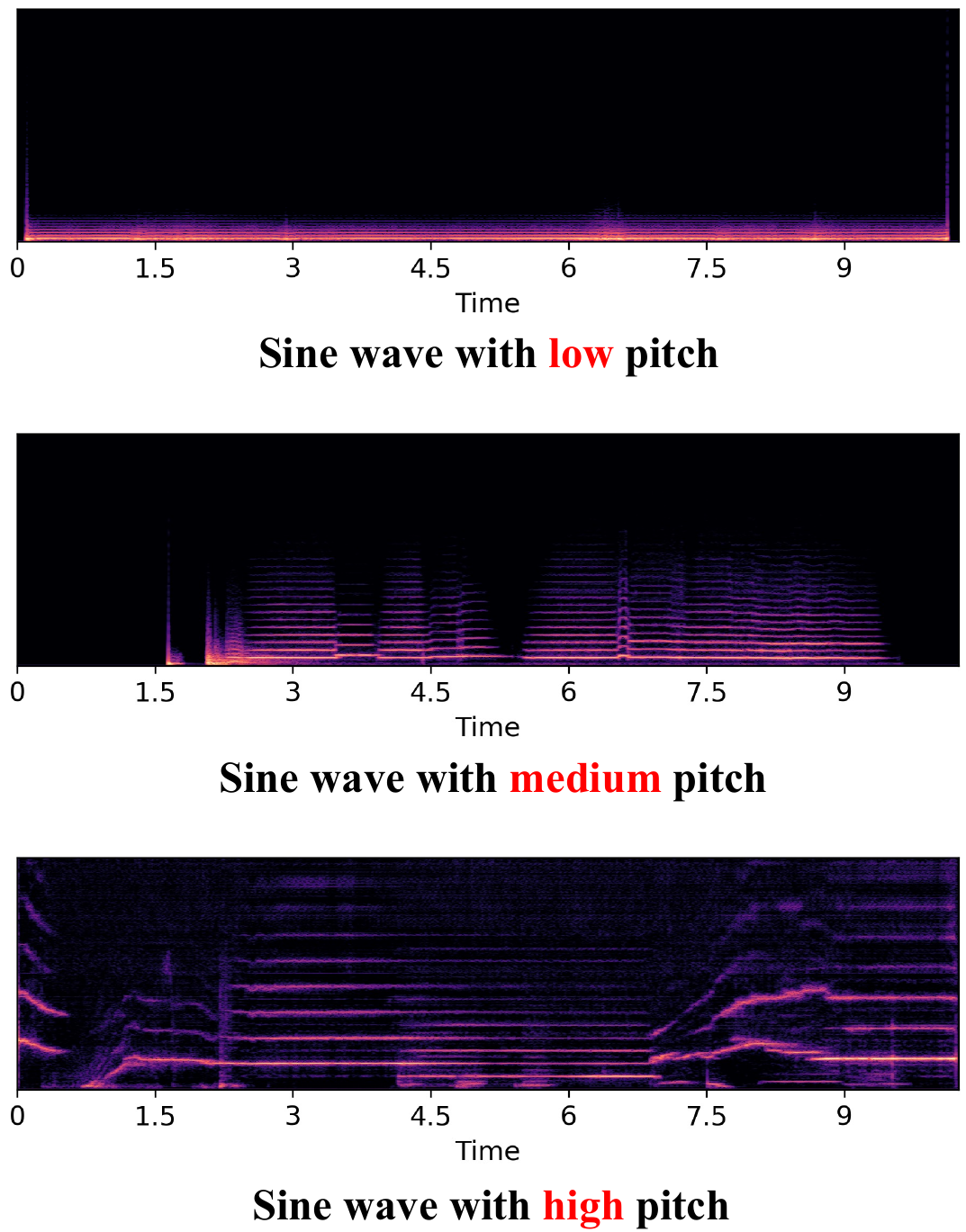}
    \caption{The examples of pitch controlling on generating samples with AudioLDM-S.}
    \label{fig:pitch}
\end{figure}

In Figure~\ref{fig:pitch}, we show the ability of AudioLDM to control the pitch of generated samples. Pitch is an important characteristic of sound effects, music and speech. Here, we set the common text information as \textbf{``Sine wave with $\cdots$ pitch''}, and input the specific text information \textbf{``low''}, \textbf{``medium''}, and \textbf{``high''}. The text-controlled pitch variation can be seen from the three generated samples.

\vspace{0.5cm}

\newpage

\textbf{Material Control}

\begin{figure}[H]
    \centering
    \includegraphics[width=1.0\linewidth]{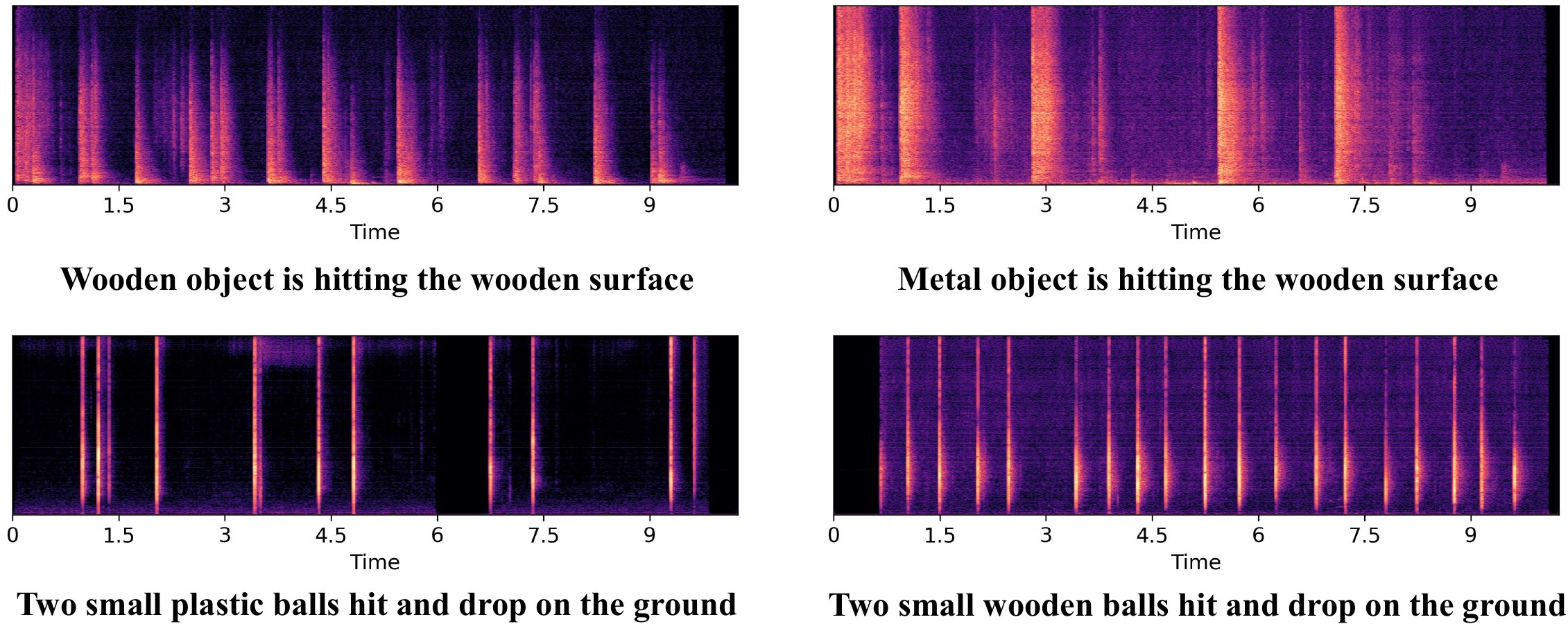}
    \caption{The examples of controlling the materials of generated audio samples with AudioLDM-S.}
    \label{fig:material}
\vspace{0.5cm}
\end{figure}

In Figure~\ref{fig:material}, we show the ability of AudioLDM to control the materials which generate audio samples. We show four samples generated by the common action \textbf{``hit''} between different materials, e.g., \textbf{wooden object} and \textbf{wooden environment}, or \textbf{metal object} and \textbf{wooden environment}. 

\vspace{0.5cm}

\vspace{0.5cm}
\textbf{Temporal Order Control}

\begin{figure}[H]
    \centering
    \includegraphics[width=1.0\linewidth]{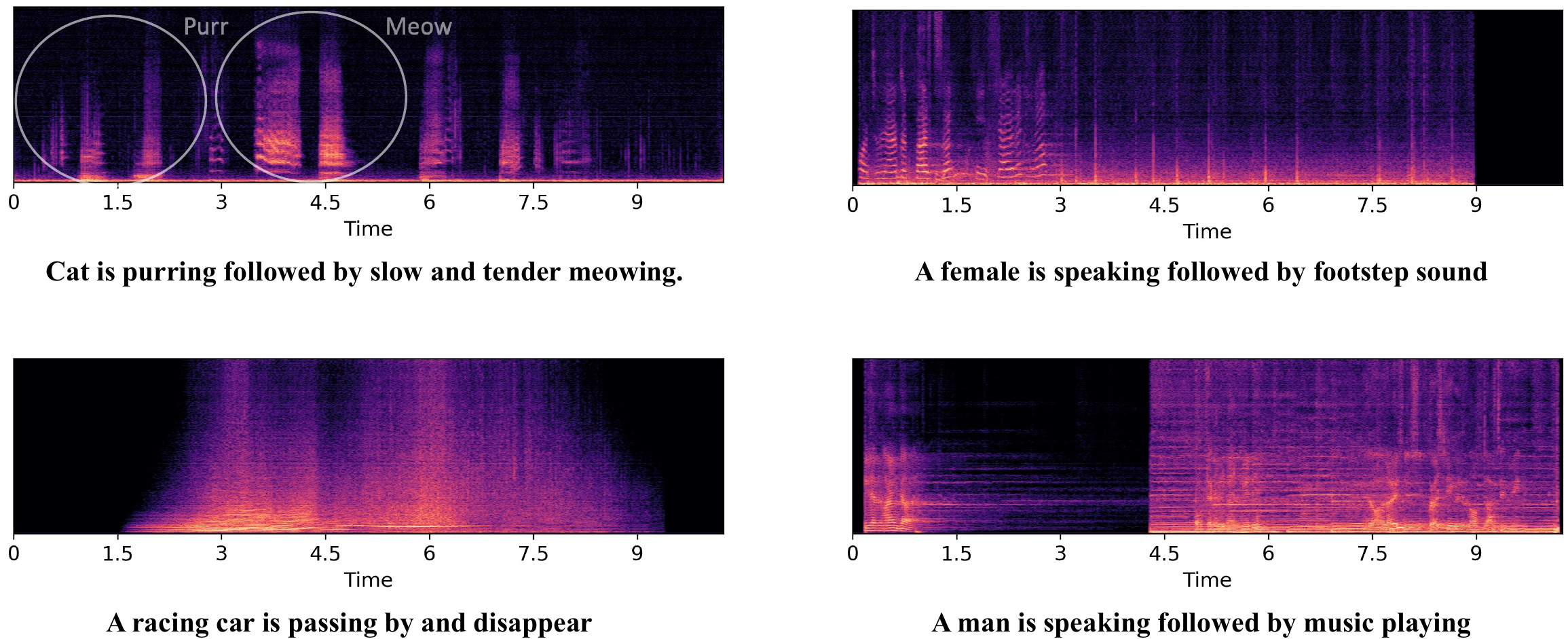}
    \caption{The examples of controlling the temporal order between generated compositional audio samples with AudioLDM-S.}
    \label{fig:temporalorder}
\vspace{0.5cm}
\end{figure}

In Figure~\ref{fig:temporalorder}, we show the ability of AudioLDM to control the temporal order between generated compositional audio signals. When the text description includes multiple sound effects, AudioLDM can generate the audio signals, and the temporal order between them is consistent with the text input.

\textbf{Text-to-Audio Generation}

\begin{figure}[H]
    \centering
    \includegraphics[width=1.0\linewidth]{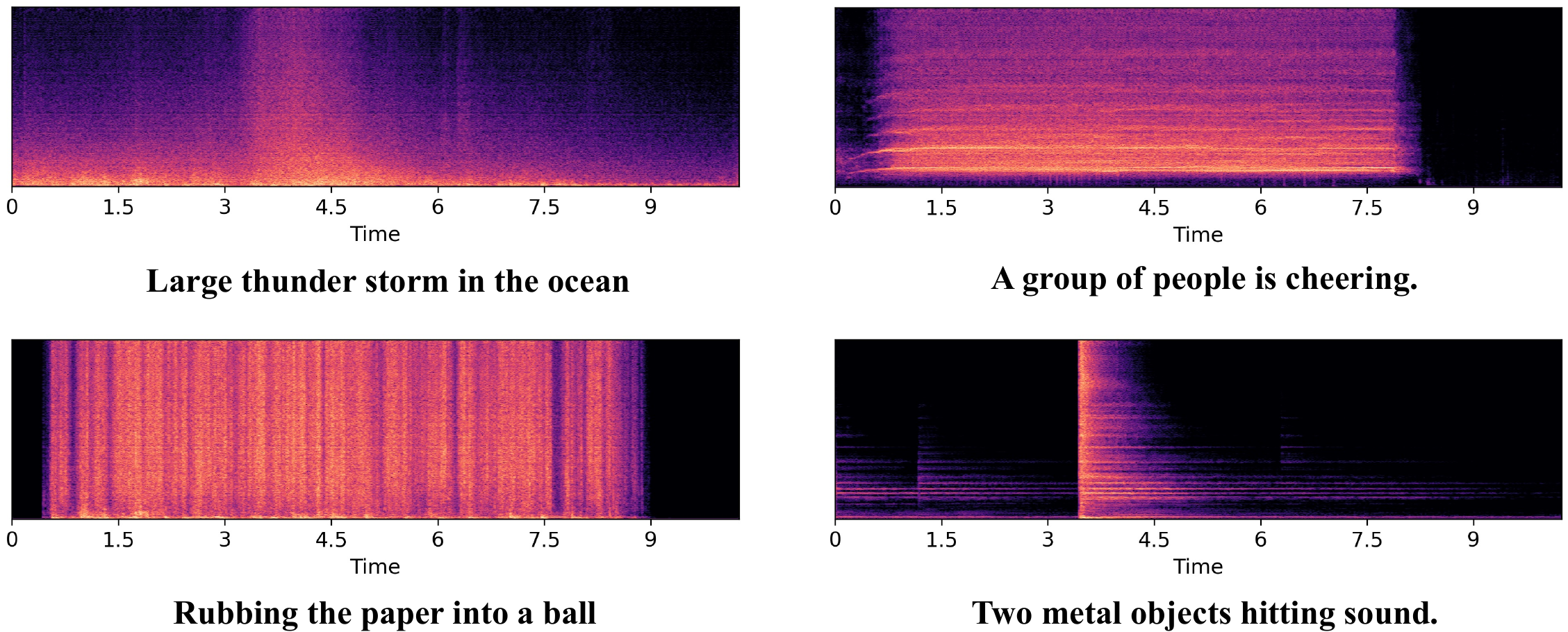}
    \caption{The examples of text-to-audio generation with AudioLDM-S.}
    \label{fig:text-to-audio}
\vspace{0.6cm}
\end{figure}

In Figure~\ref{fig:text-to-audio}, we show four text-to-audio generation results with AudioLDM-S. They include sound effects in natural environment, human speech, human activity, and sound from objects interaction.

\vspace{0.3cm}

\textbf{Novel Audio Generation}

\begin{figure}[H]
    \centering
    \includegraphics[width=1.0\linewidth]{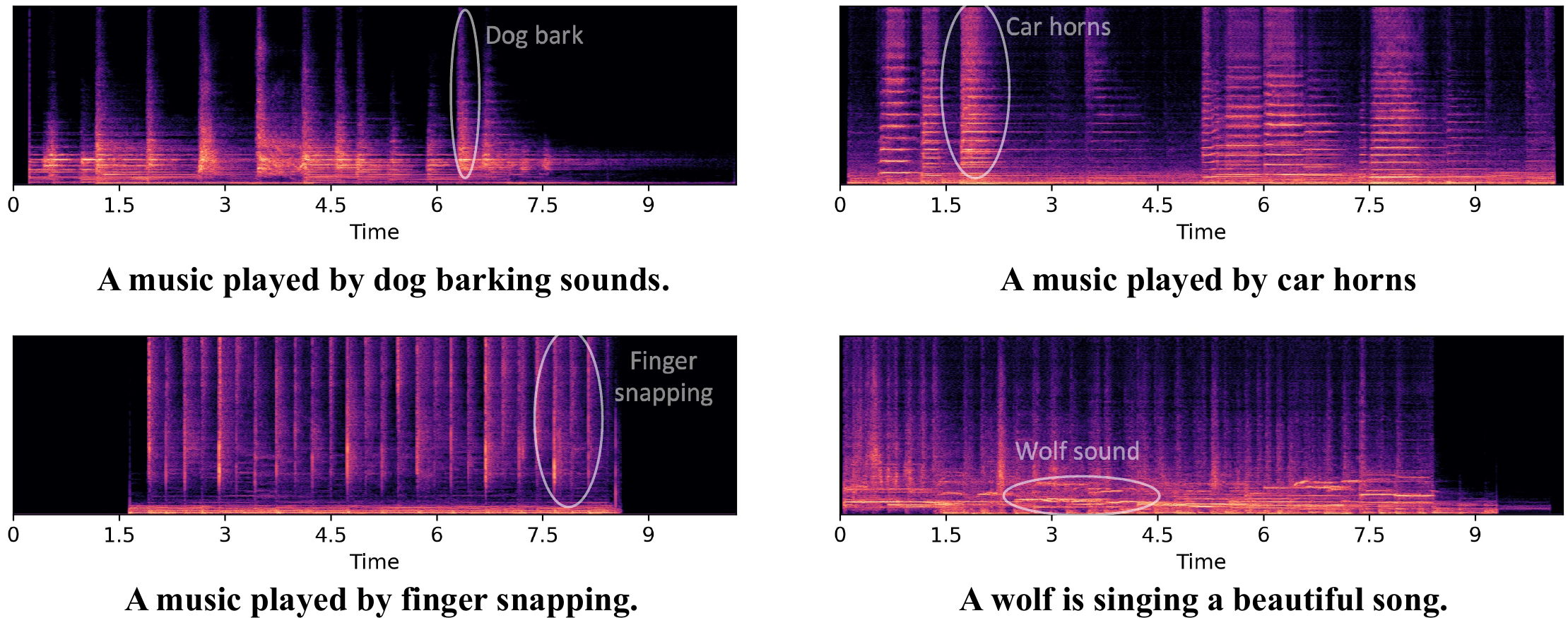}
    \caption{The examples of novel audio generation with AudioLDM-S.}
    \label{fig:novel-audio}
\end{figure}

In Figure~\ref{fig:novel-audio}, we show four novel audio samples generated with AudioLDM-S. Their text description is rarely seen, e.g., \textbf{``A wolf is singing a beautiful song.''}. We use them to exhibit the generalization ability of AudioLDM.
\vspace{0.3cm}

\newpage

\textbf{Music Generation}

\begin{figure}[H]
    \centering
    \includegraphics[width=1.0\linewidth]{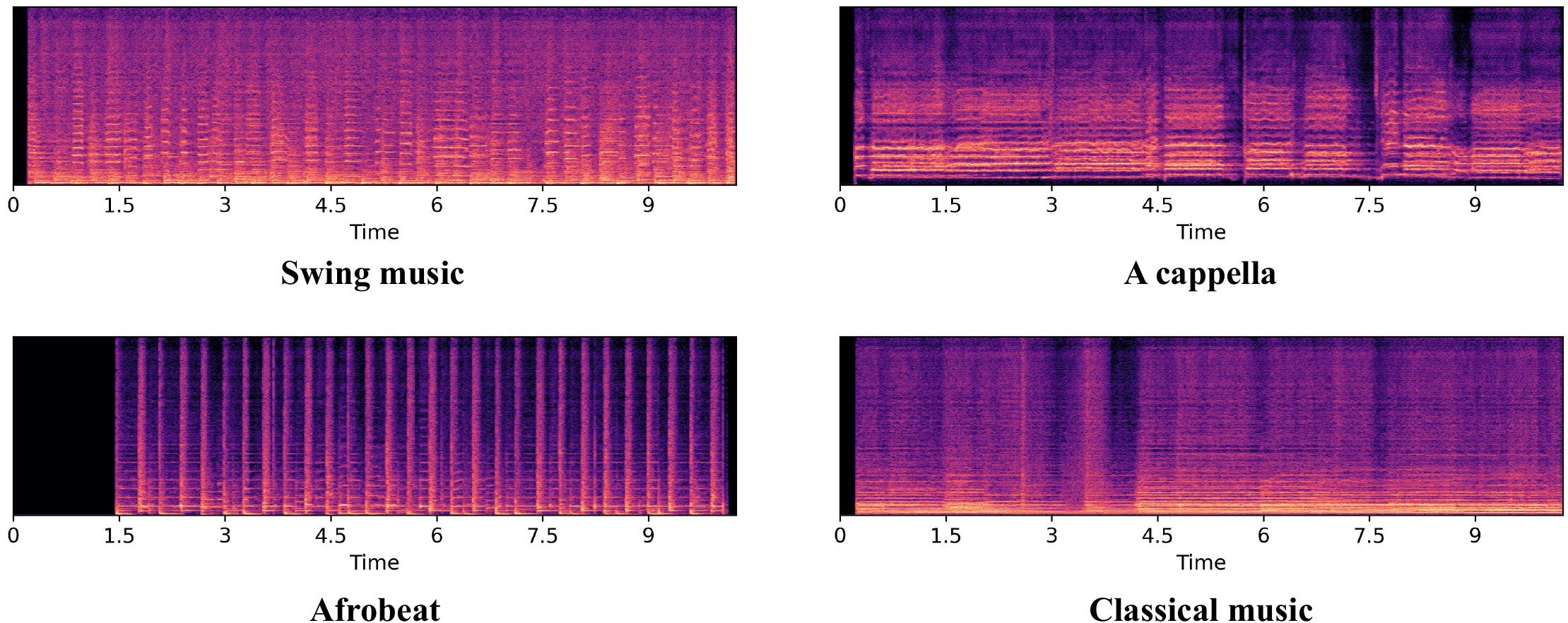}
    \caption{The examples of music generation with AudioLDM-S.}
    \label{fig:demo-audioset-music}
\end{figure}

In Figure~\ref{fig:demo-audioset-music}, we show four music samples generated with AudioLDM-S. Here, we are using the labels of AudioSet as text description for music generation, and we are able to specify the music genres of generated samples such as \textbf{Classical music}.

\newpage

\end{document}